
\catcode`@=11

\def\singlespace{\normalbaselines}
\def\oneandahalfspace{\baselineskip=1.15\normalbaselineskip plus 1pt
\lineskip=2pt\lineskiplimit=1pt}

\def\np{\vfill\eject}
\def\nl{\hfil\break}

\def\nofirstpagenoten{\nopagenumbers\footline={\ifnum\pageno>1\tenrm
\hss\folio\hss\fi}}
\def\nofirstpagenotwelve{\nopagenumbers\footline={\ifnum\pageno>1\twelverm
\hss\folio\hss\fi}}
\def\leaderfill{\leaders\hbox to 1em{\hss.\hss}\hfill}
\def\ft#1#2{{\textstyle{{#1}\over{#2}}}}
\def\frac#1/#2{\leavevmode\kern.1em
\raise.5ex\hbox{\the\scriptfont0 #1}\kern-.1em/\kern-.15em
\lower.25ex\hbox{\the\scriptfont0 #2}}
\def\sfrac#1/#2{\leavevmode\kern.1em
\raise.5ex\hbox{\the\scriptscriptfont0 #1}\kern-.1em/\kern-.15em
\lower.25ex\hbox{\the\scriptscriptfont0 #2}}

\parindent=20pt
\def\narrow{\advance\leftskip by 40pt \advance\rightskip by 40pt}

\def\AB{\bigskip
        \centerline{\bf ABSTRACT}\medskip\narrow}
\def\nonarrower{\advance\leftskip by -40pt\advance\rightskip by -40pt}
\def\AE{\bigskip\nonarrower}

\def\boxit#1{\vbox{\hrule\hbox{\vrule\kern3pt
        \vbox{\kern3pt#1\kern3pt}\kern3pt\vrule}\hrule}}

\def\gtorder{\mathrel{\raise.3ex\hbox{$>$}\mkern-14mu
             \lower0.6ex\hbox{$\sim$}}}
\def\ltorder{\mathrel{\raise.3ex\hbox{$<$}|mkern-14mu
             \lower0.6ex\hbox{\sim$}}}
\def\dalemb#1#2{{\vbox{\hrule height .#2pt
        \hbox{\vrule width.#2pt height#1pt \kern#1pt
                \vrule width.#2pt}
        \hrule height.#2pt}}}
\def\square{\mathord{\dalemb{4.9}{5}\hbox{\hskip1pt}}}

\font\fourteentt=cmtt10 scaled \magstep2
\font\fourteenbf=cmbx12 scaled \magstep1
\font\fourteenrm=cmr12 scaled \magstep1
\font\fourteeni=cmmi12 scaled \magstep1
\font\fourteenssr=cmss12 scaled \magstep1
\font\fourteenmbi=cmmib10 scaled \magstep2
\font\fourteensy=cmsy10 scaled \magstep2
\font\fourteensl=cmsl12 scaled \magstep1
\font\fourteenex=cmex10 scaled \magstep2
\font\fourteenit=cmti12 scaled \magstep1
\font\twelvett=cmtt12 \font\twelvebf=cmbx12
\font\twelverm=cmr12  \font\twelvei=cmmi12
\font\twelvessr=cmss12 \font\twelvembi=cmmib10 scaled \magstep1
\font\twelvesy=cmsy10 scaled \magstep1
\font\twelvesl=cmsl12 \font\twelveex=cmex10 scaled \magstep1
\font\twelveit=cmti12
\font\tenssr=cmss10 \font\tenmbi=cmmib10
 
 \font\ninebf=cmbx9
\font\ninerm=cmr9  \font\ninei=cmmi9
\font\ninesy=cmsy9 \font\ninessr=cmss9
\font\ninembi=cmmib10 scaled 900
\font\eightit=cmti8 \font\eightsl=cmsl8
\font\eighttt=cmtt8 \font\eightbf=cmbx8
\font\eightrm=cmr8  \font\eighti=cmmi8
\font\eightsy=cmsy8 \font\eightex=cmex10 scaled 800
\font\eightssr=cmss8 \font\eightmbi=cmmib10 scaled 800
 
\font\sevenbf=cmbx7 \font\sevenrm=cmr7 \font\seveni=cmmi7
\font\sevensy=cmsy7 
\font\sevenssr=cmss9 scaled 778 \font\sevenmbi=cmmib10 scaled 700
 
 \font\sixbf=cmbx7 scaled 875
\font\sixrm=cmr6  \font\sixi=cmmi6
\font\sixsy=cmsy6 \font\sixssr=cmss8 scaled 750
\font\sixmbi=cmmib10 scaled 600
\font\fivessr=cmss8 scaled 625  \font\fivembi=cmmib10 scaled 500

\newskip\ttglue
\newfam\ssrfam
\newfam\mbifam

\mathchardef\alpha="710B
\mathchardef\beta="710C
\mathchardef\gamma="710D
\mathchardef\delta="710E
\mathchardef\epsilon="710F
\mathchardef\zeta="7110
\mathchardef\eta="7111
\mathchardef\theta="7112
\mathchardef\iota="7113
\mathchardef\kappa="7114
\mathchardef\lambda="7115
\mathchardef\mu="7116
\mathchardef\nu="7117
\mathchardef\xi="7118
\mathchardef\pi="7119
\mathchardef\rho="711A
\mathchardef\sigma="711B
\mathchardef\tau="711C
\mathchardef\upsilon="711D
\mathchardef\phi="711E
\mathchardef\chi="711F
\mathchardef\psi="7120
\mathchardef\omega="7121
\mathchardef\varepsilon="7122
\mathchardef\vartheta="7123
\mathchardef\varpi="7124
\mathchardef\varrho="7125
\mathchardef\varsigma="7126
\mathchardef\varphi="7127
\mathchardef\partial="7140

\def\fourteenpoint{\def\rm{\fam0\fourteenrm}
\textfont0=\fourteenrm \scriptfont0=\tenrm \scriptscriptfont0=\sevenrm
\textfont1=\fourteeni \scriptfont1=\teni \scriptscriptfont1=\seveni
\textfont2=\fourteensy \scriptfont2=\tensy \scriptscriptfont2=\sevensy
\textfont3=\fourteenex \scriptfont3=\fourteenex \scriptscriptfont3=\fourteenex
\def\it{\fam\itfam\fourteenit} \textfont\itfam=\fourteenit
\def\sl{\fam\slfam\fourteensl} \textfont\slfam=\fourteensl
\def\bf{\fam\bffam\fourteenbf} \textfont\bffam=\fourteenbf
\scriptfont\bffam=\tenbf \scriptscriptfont\bffam=\sevenbf
\def\tt{\fam\ttfam\fourteentt} \textfont\ttfam=\fourteentt
\def\ssr{\fam\ssrfam\fourteenssr} \textfont\ssrfam=\fourteenssr
\scriptfont\ssrfam=\tenmbi \scriptscriptfont\ssrfam=\sevenmbi
\def\mbi{\fam\mbifam\fourteenmbi} \textfont\mbifam=\fourteenmbi
\scriptfont\mbifam=\tenmbi \scriptscriptfont\mbifam=\sevenmbi
\tt \ttglue=.5em plus .25em minus .15em
\normalbaselineskip=16pt
\bigskipamount=16pt plus5pt minus5pt
\medskipamount=8pt plus3pt minus3pt
\smallskipamount=4pt plus1pt minus1pt
\abovedisplayskip=16pt plus 4pt minus 12pt
\belowdisplayskip=16pt plus 4pt minus 12pt
\abovedisplayshortskip=0pt plus 4pt
\belowdisplayshortskip=9pt plus 4pt minus 6pt
\parskip=5pt plus 1.5pt
\twelvefoot
\setbox\strutbox=\hbox{\vrule height12pt depth5pt width0pt}
\let\sc=\tenrm
\let\big=\fourteenbig \normalbaselines\rm}
\def\fourteenbig#1{{\hbox{$\left#1\vbox to12pt{}\right.\n@space$}}
\def\square{\mathord{\dalemb{6.8}{7}\hbox{\hskip1pt}}}}

\def\twelvepoint{\def\rm{\fam0\twelverm}
\textfont0=\twelverm \scriptfont0=\ninerm \scriptscriptfont0=\sevenrm
\textfont1=\twelvei \scriptfont1=\ninei \scriptscriptfont1=\seveni
\textfont2=\twelvesy \scriptfont2=\ninesy \scriptscriptfont2=\sevensy
\textfont3=\twelveex \scriptfont3=\twelveex \scriptscriptfont3=\twelveex
\def\it{\fam\itfam\twelveit} \textfont\itfam=\twelveit
\def\sl{\fam\slfam\twelvesl} \textfont\slfam=\twelvesl
\def\bf{\fam\bffam\twelvebf} \textfont\bffam=\twelvebf
\scriptfont\bffam=\ninebf \scriptscriptfont\bffam=\sevenbf
\def\tt{\fam\ttfam\twelvett} \textfont\ttfam=\twelvett
\def\ssr{\fam\ssrfam\twelvessr} \textfont\ssrfam=\twelvessr
\scriptfont\ssrfam=\ninessr \scriptscriptfont\ssrfam=\sevenssr
\def\mbi{\fam\mbifam\twelvembi} \textfont\mbifam=\twelvembi
\scriptfont\mbifam=\ninembi \scriptscriptfont\mbifam=\sevenmbi
\tt \ttglue=.5em plus .25em minus .15em
\normalbaselineskip=14pt
\bigskipamount=14pt plus4pt minus4pt
\medskipamount=7pt plus2pt minus2pt
\abovedisplayskip=14pt plus 3pt minus 10pt
\belowdisplayskip=14pt plus 3pt minus 10pt
\abovedisplayshortskip=0pt plus 3pt
\belowdisplayshortskip=8pt plus 3pt minus 5pt
\parskip=3pt plus 1.5pt
\tenfoot
\setbox\strutbox=\hbox{\vrule height10pt depth4pt width0pt}
\let\sc=\ninerm
\let\big=\twelvebig \normalbaselines\rm}
\def\twelvebig#1{{\hbox{$\left#1\vbox to10pt{}\right.\n@space$}}
\def\square{\mathord{\dalemb{5.9}{6}\hbox{\hskip1pt}}}}

\def\tenpoint{\def\rm{\fam0\tenrm}
\textfont0=\tenrm \scriptfont0=\sevenrm \scriptscriptfont0=\fiverm
\textfont1=\teni \scriptfont1=\seveni \scriptscriptfont1=\fivei
\textfont2=\tensy \scriptfont2=\sevensy \scriptscriptfont2=\fivesy
\textfont3=\tenex \scriptfont3=\tenex \scriptscriptfont3=\tenex
\def\it{\fam\itfam\tenit} \textfont\itfam=\tenit
\def\sl{\fam\slfam\tensl} \textfont\slfam=\tensl
\def\bf{\fam\bffam\tenbf} \textfont\bffam=\tenbf
\scriptfont\bffam=\sevenbf \scriptscriptfont\bffam=\fivebf
\def\tt{\fam\ttfam\tentt} \textfont\ttfam=\tentt
\def\ssr{\fam\ssrfam\tenssr} \textfont\ssrfam=\tenssr
\scriptfont\ssrfam=\sevenssr \scriptscriptfont\ssrfam=\fivessr
\def\mbi{\fam\mbifam\tenmbi} \textfont\mbifam=\tenmbi
\scriptfont\mbifam=\sevenmbi \scriptscriptfont\mbifam=\fivembi
\tt \ttglue=.5em plus .25em minus .15em
\normalbaselineskip=12pt
\bigskipamount=12pt plus4pt minus4pt
\medskipamount=6pt plus2pt minus2pt
\abovedisplayskip=12pt plus 3pt minus 9pt
\belowdisplayskip=12pt plus 3pt minus 9pt
\abovedisplayshortskip=0pt plus 3pt
\belowdisplayshortskip=7pt plus 3pt minus 4pt
\parskip=0.0pt plus 1.0pt
\eightfoot
\setbox\strutbox=\hbox{\vrule height8.5pt depth3.5pt width0pt}
\let\sc=\eightrm
\let\big=\tenbig \normalbaselines\rm}
\def\tenbig#1{{\hbox{$\left#1\vbox to8.5pt{}\right.\n@space$}}
\def\square{\mathord{\dalemb{4.9}{5}\hbox{\hskip1pt}}}}

\def\eightpoint{\def\rm{\fam0\eightrm}
\textfont0=\eightrm \scriptfont0=\sixrm \scriptscriptfont0=\fiverm
\textfont1=\eighti \scriptfont1=\sixi \scriptscriptfont1=\fivei
\textfont2=\eightsy \scriptfont2=\sixsy \scriptscriptfont2=\fivesy
\textfont3=\eightex \scriptfont3=\eightex \scriptscriptfont3=\eightex
\def\it{\fam\itfam\eightit} \textfont\itfam=\eightit
\def\sl{\fam\slfam\eightsl} \textfont\slfam=\eightsl
\def\bf{\fam\bffam\eightbf} \textfont\bffam=\eightbf
\scriptfont\bffam=\sixbf \scriptscriptfont\bffam=\fivebf
\def\tt{\fam\ttfam\eighttt} \textfont\ttfam=\eighttt
\def\ssr{\fam\ssrfam\eightssr} \textfont\ssrfam=\eightssr
\scriptfont\ssrfam=\sixssr \scriptscriptfont\ssrfam=\fivessr
\def\mbi{\fam\mbifam\eightmbi} \textfont\mbifam=\eightmbi
\scriptfont\mbifam=\sixmbi \scriptscriptfont\mbifam=\fivembi
\tt \ttglue=.5em plus .25em minus .15em
\normalbaselineskip=9pt
\bigskipamount=9pt plus3pt minus3pt
\medskipamount=5pt plus2pt minus2pt
\abovedisplayskip=9pt plus 3pt minus 9pt
\belowdisplayskip=9pt plus 3pt minus 9pt
\abovedisplayshortskip=0pt plus 3pt
\belowdisplayshortskip=5pt plus 3pt minus 4pt
\parskip=0.0pt plus 1.0pt
\setbox\strutbox=\hbox{\vrule height8.5pt depth3.5pt width0pt}
\let\sc=\sixrm
\let\big=\eightbig \normalbaselines\rm}
\def\eightbig#1{{\hbox{$\left#1\vbox to6.5pt{}\right.\n@space$}}
\def\square{\mathord{\dalemb{3.9}{4}\hbox{\hskip1pt}}}}

\def\vfootnote#1{\insert\footins\bgroup\footsuite
    \interlinepenalty=\interfootnotelinepenalty
    \splittopskip=\ht\strutbox
    \splitmaxdepth=\dp\strutbox \floatingpenalty=20000
    \leftskip=0pt \rightskip=0pt \spaceskip=0pt \xspaceskip=0pt
    \textindent{#1}\footstrut\futurelet\next\fo@t}
\def\hangfootnote#1{\edef\@sf{\spacefactor\the\spacefactor}#1\@sf
    \insert\footins\bgroup\footsuite
    \let\par=\endgraf
    \interlinepenalty=\interfootnotelinepenalty
    \splittopskip=\ht\strutbox
    \splitmaxdepth=\dp\strutbox \floatingpenalty=20000
    \leftskip=0pt \rightskip=0pt \spaceskip=0pt \xspaceskip=0pt
    \smallskip\item{#1}\bgroup\strut\aftergroup\@foot\let\next}
\def\footsuite{}
\def\twelvefoot{\def\footsuite{\twelvepoint}}
\def\tenfoot{\def\footsuite{\tenpoint}}
\def\eightfoot{\def\footsuite{\eightpoint}}
\catcode`@=12

\def\Re{{\rm\rlap I\mkern3mu R}}
\def\im{{\rm i}}

\def\ft#1#2{{\textstyle{{#1}\over{#2}}}}
\def\frac#1#2{\leavevmode\kern.1em
\raise.5ex\hbox{\the\scriptfont0 #1}\kern-.1em/\kern-.15em
\lower.25ex\hbox{\the\scriptfont0 #2}}
\def\sfrac#1#2{\leavevmode\kern.1em
\raise.5ex\hbox{\the\scriptscriptfont0 #1}\kern-.1em/\kern-.15em
\lower.25ex\hbox{\the\scriptscriptfont0 #2}}

\def\cramp{\medmuskip = 2mu plus 1mu minus 2mu}

\def\crampest{\medmuskip = 1mu plus 1mu minus 1mu}
\def\uncramp{\medmuskip = 4mu plus 2mu minus 4mu}
\twelvepoint
\nofirstpagenotwelve
\oneandahalfspace

\rightline{CTP TAMU--18/94}
\rightline {Imperial/TP/93-94/31}
\rightline{hep-th/9404170}
\rightline{April 1994}
\vskip 3truecm

\centerline{\bf Canonical BRST Quantisation of Worldsheet Gravities}

\vskip 1truecm

\centerline{R. Mohayee$^1$, C.N. Pope$^2$\footnote{$^{\star}$}{\tenfoot
Supported in part by the U.S. Department of Energy, under grant
DE-FG05-91-ER40633}, K.S. Stelle$^1$ and K.-W. Xu$^2$}
\vskip 1truecm
\centerline{\it $^{1}$ The Blackett Laboratory,}
\centerline{\it Imperial College, London SW7 2BZ, England}
\bigskip
\centerline{\it $^{2}$ Center for Theoretical Physics, Department of Physics}
\centerline{\it Texas A\&M University, College Station, TX 77843--4242, U.S.A.}
\vskip 1truecm

\singlespace\AB
   We reformulate the BRST quantisation of chiral Virasoro and $W_3$
worldsheet gravities. Our approach follows directly the classic BRST
formulation of Yang-Mills theory in employing a derivative gauge condition
instead of the conventional conformal gauge condition, supplemented by an
introduction of momenta in order to put the ghost action back into first-order
form. The consequence of these simple changes is a considerable simplification
of the BRST formulation, the evaluation of anomalies and the expression
of Wess-Zumino consistency conditions. In particular, the transformation rules
of all fields now constitute a canonical transformation generated by the
BRST operator $Q$, and we obtain in this reformulation a new result that the
anomaly in the BRST Ward identity is obtained by application of the anomalous
operator $Q^2$, calculated using operator products, to the gauge fermion.
\AE\oneandahalfspace
\np
\noindent{\bf 1. Introduction}
\bigskip

     The BRST formalism has proven to be the most powerful approach to the
quantisation of string theories. Indeed, the full spectrum of low-dimensional
string and $W$-string theories can only be properly derived in the BRST
formalism [1,2]. The appropriateness of the BRST formalism is owed to the
control it gives in the handling of anomalies in world-sheet chiral algebras.
In the case of the non-critical bosonic string, the  worldsheet anomaly gives
rise to a propagating Liouville mode whose presence renders the worldsheet
``gravity'' non-trivial. Analogous anomalies in the $W_3$ string give rise to a
worldsheet $W_3$ gravity described by an $A_2$ Toda theory [3].

     There exist two basic approaches to the treatment of such
anomaly-induced dynamics. The standard approach of Liouville and Toda
gravities is to anticipate the occurrence of the anomalies already at the
classical level by introducing classically-decoupling compensating fields.
These fields maintain the full worldsheet symmetries at the quantum level by
construction, but the anomalies arise in this approach through an anomalous
quantum-level coupling of the compensators to the other fields of the theory.
The second approach is to extract the anomalous quantum dynamics directly from
the the anomalous Ward identities of the worldsheet symmetries. In this latter
approach, non-trivial correlation functions arise at the quantum level,
revealing in some cases hidden quantum symmetries such as the $SL(2,\Re)$
symmetry found by Polyakov for the bosonic string [4], or the $SL(\infty,\Re)$
symmetry found in worldsheet $W_\infty$ gravity [5,6] (which becomes a
$GL(\infty,\Re)$ symmetry for $W_{1+\infty}$ gravity [6].) Because it reveals
hidden symmetries, the approach of extracting dynamics from anomalous Ward
identities is clearly of great importance for the non-critical theories.
The symmetries so found are reminiscent of the underlying $A_N$ symmetries for
the Liouville or Toda theories in the first approach, but the precise
relationship between these symmetries still needs clarification.

     Analysis of the anomalous Ward identities for nonlinear chiral worldsheet
algebras such as $W_3$ is made more difficult by the complexity and
off-diagonal
nature of the anomalies. Attempts in [7] to extend the
approach of [4--6] to the $W_N$ gravity case ran into the difficulty that a
consistent set of conditions to impose on the background gauge fields to
eliminate the anomalies could not be derived owing to their off-diagonal
structure.  These difficulties are presumably related to our imperfect
understanding of $W_3$ geometry.

     In this paper, we present a reformulation of the BRST quantisation
procedure for worldsheet gravities and the derivation of anomalous Ward
identities. We hope that this will prove useful for understanding the dynamics
of non-critical worldsheet gravities. In our reformulation, we shall first have
to choose a fully acceptable gauge condition. It is well-known that the
standard conformal gauge condition employed in string theory [8] is not really
an acceptable choice, because in making it one looses the Virasoro constraints
as field equations. Acceptable gauges may be defined as gauge conditions that
can be imposed either prior to or after varying the action in order to obtain
the classical equations of motion. The easiest way to make an acceptable gauge
choice is to choose a derivative gauge.\footnote{$^{\ast}$}{\tenfoot Earlier
discussions of derivative gauges in string theory, such as the harmonic gauge,
may be found in Refs [9].} The point is most easily expressed
by comparison to Maxwell electrodynamics. In Maxwell theory, which has a local
$U(1)$ gauge symmetry, the Lagrangian density is
$$
{\cal L}=-{1\over 4} F_{\mu \nu }F^{\mu \nu}
=-{1\over 4} \Big[ 2 F_{0i}F^{0i}+
F_{ij}F^{ij} \Big],
\eqno (1.1)
$$
where the signature is $(-1,1,1,1)$, $F_{\mu \nu}={\partial }_{\mu}
A_{\nu} - {\partial }_{\nu} A_{\mu} $, $A_{\mu}$ is the gauge
field, and $\mu=0,i$. The canonical momenta are defined as
$$
{\pi }_{\mu} = {{\partial }{\cal L}\over \partial [{\partial }_0 A^{\mu}]},
\eqno (1.2)
$$
namely ${\pi }_0=0$, ${\pi }_i=F_{0i}=-E_i$, where $E_i$ is the
electric field.  The equal-time Poisson Bracket (PB) is
$$
\Big\{{\pi }_i(\vec x,t) , A_j(\vec y,t)\Big\}_{PB} =
\delta_{ij}\, {\delta }(\vec x - \vec y).
\eqno (1.3)
$$
The Hamiltonian density is
$$
{\cal H}= {1\over 4} F_{ij}F^{ij}+{1\over 2} {\pi}_i\,
{\pi}^i - A_0\, {\partial }_i {\pi }^i ,
\eqno (1.4)
$$
and the canonical action for evolution from $t_0$ to $t_f$ is
$$
I={\int }^{t_f}_{t_0} dt \int d^3x\,\big(\pi_i\,\partial_0 A^i - {\cal
H}\big).\eqno(1.5)
$$
The equation of motion obtained by varying the Lagrange multiplier
$A_0$ in the action (1.5) is
$$
{\partial }_i{\pi }^i(\vec x,t)=0.
\eqno (1.6)
$$
Had we insisted on setting $A_0=0$ prior to varying the action we
would have lost the Gauss' law constraint $\nabla \cdot E =0$ as an
equation of motion. Consequently, we need to choose a different gauge
such as the Lorentz gauge ${\partial }^{\mu} A_{\mu} =0$ to fix the
$U(1)$ gauge symmetry.

  Another way of saying this [10] is to suppose that $A_0=\epsilon (\vec x,t)
\not = 0$, and to try to make a gauge transformation so as to move into $A_0=0$
gauge. As the gauge transformation is
$$
{\delta }A_0={\dot \lambda }(\vec x,t),
\eqno (1.7)
$$
one would have to solve the first order differential equation
$$
{\dot \lambda }(\vec x,t)={\epsilon }(\vec x,t).
\eqno (1.8)
$$
Now, recall that one obtains the equations of motion by varying the
fields in the action (1.5) subject to the endpoint conditions
$\delta A_{\mu}(t_0)=\delta A_{\mu}(t_f)=0$. This determines the
Euler-Lagrange evolution of the fields between the initial and final field
configurations $A_{\mu}(\vec x,t_0)$ and $A_{\mu}(\vec x,t_f)$. The gauge
symmetries for this variational problem are defined to be those
transformations that leave the action (1.5) invariant, with the same fixed
initial and final times as chosen in varying the fields. Making a gauge
transformation $\delta A_\mu=\partial_\mu\lambda$, $\delta\pi_i=0$ on the
fields appearing in (1.5), we find
$$
\delta I = \int d^3x\,\lambda(\vec
x,t)\,\partial_i\pi^i{\textstyle\Big|^{t_f}_{t_0}}.\eqno(1.9)
$$
Thus, requiring invariance of (1.5) for general field configurations at
$t_0$ and $t_f$ requires
$$
{\lambda }(\vec x, t_0)={\lambda }(\vec x, t_f)=0.
\eqno (1.10)
$$
For the first-order differential equation (1.8), imposing the two boundary
conditions on $\lambda$ overdetermines the problem, yielding no solution.
Thus, the $A_0=0$ condition is not one that can actually be achieved for
the canonical action (1.5) by a gauge transformation starting from a general
field configuration.

  By contrast, in the covariant gauge ${\partial }_{\mu} A^{\mu} =0$, the
differential equation for the transformation parameter becomes of second order
$$
{\ddot \lambda }(\vec x,t)={\epsilon }(\vec x,t),
\eqno (1.11)
$$
and allows the imposition of two boundary conditions, so we can actually find a
solution to move into such a gauge.

   In this paper, we shall concentrate on chiral worldsheet gravities and shall
study the BRST formulation and gauge fixing of a single copy of a chiral
worldsheet Virasoro or $W_3$ algebra. One may either view these chiral
worldsheet gravities as theories defined in their own right, or may take the
point of view that they arise from some theory, such as a bosonic string or
$W_3$-string theory, that has undergone a preliminary stage of gauge fixing
that includes the condition
$$
\gamma_{ij}=\pmatrix{0&1\cr1&h} , \eqno(1.12)
$$
in complex light-cone variables $z,\bar z$. This gauge condition leaves
unfixed a residual gauged chiral algebra, with a remaining gauge field $h$
(together with a spin-3 field $B$ in the $W_3$ case). Whether or not the gauge
condition (1.12) may legitimately be imposed in the sense
of Ref.\ [10] in string theory, or in view of hermiticity
requirements, is itself an interesting question. However those questions will
lie outside the scope of the present paper, where we shall take the chiral
worldsheet gravities as our starting points. We shall be more careful in our
discussion of the the final stage of gauge fixing, where we shall replace the
conventional conformal gauge conditions $h=h_{\rm back}$, $B=B_{\rm back}$
by derivative gauges such as $\bar\partial h=\bar\partial B=0$.

  We shall derive the BRST charge for our construction from the BRST
transformations using Noether's theorem. It is well-known that the
transformation rule for the spin-2 gauge field $h$ cannot ordinarily be
obtained
directly from the standard BRST charge in the conventional formulation,
since the standard BRST charge contains nothing with non-vanishing
commutator or anticommutator brackets with $h$. This asymmetrical treatment of
the gauge fields is compounded in the $W_3$ case by the fact that the BRST
transformations of $h$ and the matter fields $\varphi^i$ do not form a closed
nilpotent algebra unless one uses the classical $h$ and $\varphi^i$ equations
of motion [11]. Quantisation in such a situation may be handled within the
context of Batalin-Vilkovisky (BV) quantisation [12], but at the price, in the
$W_3$ case [13], of a significant increase in complexity with respect to the
quantisation procedure that we shall present.

  In the reformulated BRST approach of this paper, we shall impose
derivative gauge conditions and shall replace the resulting second-order
ghost-sector actions with first-order actions after an introduction of momenta
as auxiliary variables. The resulting gauge-fixed Virasoro and  $W_3$ theories
will be shown to be classically invariant under a set of BRST transformations
of all fields that can now be obtained as a
canonical\footnote{$^{\dag}$}{\tenfoot Related work on he canonical
approach to the BV quantisation of string theory in derivative gauges may
be found in Ref.\ [14], where the gauge independence of the BRST charge $Q$
is also obtained.} transformation because the associated Noether charge
$Q$ now properly generates the full set of BRST transformations of all
fields, including the gauge fields. In deriving the BRST transformations
generated by the chiral charge $Q$, we shall be treating the complex
Euclidean worldsheet coordinates $z$ and $\bar z$ as
independent\footnote{$^{\ddag}$}{\tenfoot This independence is of course
naturally obtained in a Minkowski-space formulation, where
$\sigma\pm\tau$ are truely independent. We shall be using the
Euclidean-space formulation to facilitate later comparison with quantum
operator-product calculations.} and we shall treat the $\bar\partial$
derivative as the ``evolution'' derivative in the definition of momenta.
As usual, however, the non-invariance of the path-integral measure for the
partition function under Weyl transformations generally modifies this story
at the quantum level by the presence of BRST anomalies.

  A very natural but apparently new result that comes out of our reformulation
of the BRST quantisation procedure is the precise relation between two notions
of anomalies that one may encounter in the literature on worldsheet gravities.
In conformal field theory, the notion of an anomaly is concerned with a
violation of the classical BRST algebra, {\it i.e.}\ with a loss of
nilpotence for $Q$ at the quantum level, and this is evaluated by taking a
fully-contracted operator product $Q^2$, yielding a local but non-vanishing
anomalous result.  In ordinary field theory, the notion of an anomaly is
concerned with the violation of the BRST Ward identity for the effective action
$\Gamma$, {\it i.e.}\ the ``master equation'' of the Batalin-Vilkovisky (BV)
formalism [12], and is evaluated by considering the 1PI diagrams, leading to
an anomalous Ward identity. We shall see that the local functional expressing
the anomaly in the BRST Ward identity is given by the operator product of the
anomalous local operator $Q^2$ with the ``gauge fermion'' $\Psi$ of our
reformulated gauge-fixed theory.

  We shall calculate the chiral Virasoro anomaly at one loop and the
local anomalies of chiral $W_3$ gravity at one and two loops. The Virasoro
anomaly that we shall find is the same as the one derived by Polyakov by a
straightforward variation of the effective action. The situation for $W_3$ is
more subtle, and we shall obtain an anomaly that differs from the one given in
[7] both in the values of coefficients and also in the occurrence of new terms.
We shall check the correctness of our expressions for the Virasoro and $W_3$
anomalies by verifying that the Wess-Zumino consistency conditions are
satisfied in each case. Whether our results for the $W_3$ anomalies are
actually
in conflict with those of Ref.\ [7] or not remains to be determined, however,
since the analysis of Ref.\ [7] followed a different procedure of treating the
gauge fields purely as external fields without gauge fixing or introduction of
ghosts, and sought in this way to derive the dynamics of ``induced gravity.''

  The paper is organised as follows. In the next section, we shall quantise
chiral Virasoro gravity with a derivative gauge condition. We shall derive the
BRST charge, calculate the Virasoro anomaly and check that the Wess-Zumino
consistency condition is satisfied. In section three, we shall review the
conventional BV quantisation of
$W_3$ gravity starting from the results of [15]. This has also recently been
discussed in greater detail in [13]; the purpose of our review will be to
set the stage for our reformulated treatment of the $W_3$ case, which will be
given in section four. In the concluding section, we shall present the
relation between anomalies in the quantum $Q^2$ algebra and in the BRST Ward
identities that can be abstracted from our reformulated quantisation procedure.

\bigskip
\noindent{\bf 2. BRST Quantisation of Virasoro gravity}
\bigskip

  In this section, we will develop our point of view on BRST quantisation by
focusing on the Virasoro case. For comparison, we start by reviewing the
conventional conformal-gauge BRST quantisation of worldsheet Virasoro gravity
[8].

\bigskip
\noindent{\it Conventional BRST quantisation}
\medskip

  The chiral Virasoro gravity action in the preliminary gauge (1.12) is
$$
I_0={1\over \pi} \int d^2z \Big(-{1\over 2} {\bar \partial } {\varphi }^i
\,{\partial } {\varphi }^i + {1\over 2} h\,
{\partial } {\varphi }^i \,{\partial } {\varphi }^i \Big) ,
\eqno (2.1)
$$
where the ${\varphi }^i$ ( $ i=0,1, \dots, D-1$, and D is the space time
dimension) are a set of matter fields and $h$ is the remaining unfixed
component of the two-dimensional metric. This action is invariant under the
following transformations:
$$
\eqalignno{&  \delta \varphi =\varepsilon\,
 {\partial {\varphi }^i} , &(2.2a) \cr
& \delta h ={\bar \partial } {\varepsilon } + {\varepsilon }\, \partial h -
{\partial \varepsilon }\, h .  &(2.2b) \cr }
$$
The standard conformal-gauge way to fix this gauge freedom is to set
$h=h_{back}$. One then obtains the gauge-fixed action
$$
I={1\over \pi} \int d^2z \Big( -{1\over 2}
{\bar\partial{\varphi }^i }\,\partial {\varphi }^i
- b\, {\bar\partial {c} } + {\pi }_h (h-h_{back}) - h(T_{\rm mat}+T_{\rm gh})
\Big).
\eqno(2.3)
$$
This action has the following BRST symmetry
$$
\eqalignno{&  \delta {\varphi }^i=c\, {\partial {\varphi }^i} ,  &(2.4a) \cr
& \delta h ={\bar \partial }c + c\, \partial h - {\partial c}\, h , &(2.4b) \cr
& \delta c = c \,\partial c  ,  &(2.4c) \cr
& \delta b = {\pi }_h  , &(2.4d) \cr
& \delta {\pi }_h = 0 . &(2.4e) \cr }
$$
where ${\pi }_h $ is an auxiliary field and $c$, $b$ denote ghost and
anti-ghost fields, satisfying standard OPE relations. $T_{\rm mat}$
and $T_{\rm gh}$ are the energy-momentum tensors for the matter fields and
ghost fields respectively, and are given explicitly by
$$
\eqalignno{ T_{\rm mat}&=-{1\over 2} \partial {\varphi }^i \,\partial {\varphi
}^i ,
 &(2.5a) \cr
T_{\rm gh}&= -2 b\, \partial {c}
- {\partial {b} }\, c . &(2.5b) \cr }
$$
The operator products of both $T_{\rm mat}$ and $T_{\rm gh}$ close to form
the OPE Virasoro algebra
$$
\hbar^{-1}T(z)T(w) \sim { {\partial T }\over {z-w}} +
  {{2T}\over {(z-w)^2}}
+\hbar{{ 1\over 2 } C\over {(z-w)^4}},
\eqno (2.6)
$$
where $C$ is the central charge.

  From the action (2.3) and the BRST transformations (2.4),
one may construct the conserved charge related to this symmetry by Noether's
theorem. In this way, one obtains the standard BRST charge for chiral Virasoro
gravity,
$$
  Q= \int dz\, c\,\Big(T_{\rm mat}+{1\over 2}T_{\rm gh} \Big) .
\eqno(2.7)
$$

  At this point, we encounter the difficulties with this standard procedure. In
a properly-posed canonical formalism, the BRST charge should act as the
generator of all of the transformations from which it was originally derived,
{\it i.e.}\ one would like to have
$$
\delta {\phi }^i = \Big\{ Q, {\phi }^i \Big\},
\eqno (2.8)
$$
where the $\{\cdot,\cdot\}$ bracket is realised either classically as a Poisson
bracket/antibracket or quantum-mechanically as a commutator/anticommutator.
Now, this works as expected for the fields ${\varphi }^i$ and
$c$. If one substitutes the equations of motion, it also works for $b$ and
${\pi }_h$. But it can never work for the $h$ field, since nothing in (2.7)
has any nontrivial commutation properties with $h$. The BRST transformation of
the $h$ field can only be derived from the requirement of invariance of the
action (2.3) under the BRST transformations. This is an analogue of the
situation in $A_0=0$ gauge in Maxwell theory. Similarly to this Maxwell
gauge, where Gauss' law is lost as an equation of motion, in the Virasoro
gravity case one has to remember separately to impose the Virasoro
constraints by hand, since the
$h$ field equation following from (2.3) implies only that $T_{\rm mat}+T_{\rm
gh}=\pi_h$. We shall see that all of these problems are resolved
naturally in our revised BRST quantisation procedure.
\bigskip
\noindent{\it Derivative-gauge BRST quantisation}
\medskip

  We return to the action (2.1) in the preliminary gauge choice (1.12), but
now complete the gauge fixing by choosing the derivative gauge condition ${\bar
\partial }h=0$. The gauge-fixed action then becomes
$$
\eqalign{I &={1\over \pi} \int d^2z\, {\cal L} \cr
&= {1\over \pi} \int d^2z \Big( -{1\over 2}
{\bar\partial{\varphi }^i }\,\partial {\varphi }^i
- h\, T_{\rm mat} + {\pi }_h\, {\bar \partial }h -
b\, {\bar \partial }({\bar \partial }c +
c\, \partial h - {\partial c}\, h ) \Big). \cr}\eqno(2.9)
$$
As a result of the derivative gauge condition, the ghost action is now of
second order in $\bar\partial$ derivatives. For comparison with
conformal-field-theory operator products and also to allow us to use the
canonical formalism, we next introduce auxiliary fields in order to put the
ghost sector into first-order form. From the action (2.9), one sees that the
fields $c$ and $b$ are no longer conjugates, so we need to define conjugate
momenta
$$
\eqalign{{\pi }_c &={{\partial }{\cal L}\over \partial {\bar \partial }c}=
-{\bar \partial }b , \cr
{\pi }_b &={{\partial }{\cal L}\over \partial {\bar \partial }b}=
{\bar \partial }c + c \,\partial h - {\partial c}\, h . \cr}\eqno (2.10)
$$
As noted in the introduction, throughout this paper we shall be treating
the $\bar\partial$ derivative as the ``evolution'' derivative in the
canonical formulation of the chiral theories we consider. We can rewrite
the second-order action (2.9) in first-order form as
$$
I = {1\over \pi} \int d^2z \Big( -{1\over 2}
{\bar\partial{\varphi }^i }\,\partial {\varphi }^i
+ {\pi }_h\, {\bar \partial }h - {\pi }_b\, {\bar \partial }b
-{\pi }_c \,{\bar \partial }c -{\pi }_b\, {\pi }_c
-h\, ( T_{\rm mat} + T_{\rm gh} )
\Big).\eqno(2.11)
$$
 From the path-integral generating functional derived from this action, we get
the following OPE relations
$$
\eqalign{ \partial {\varphi }^i (z) \partial {\varphi }^i (w) \sim {- {\hbar
}\over (z-w)^2}\ \ \ \ \ \ \ \ &\ \ \ \ \ \ \  {\pi }_h (z) h (w) \sim {{\hbar
}\over z-w } \cr c(z){\pi }_c (w) \sim {{\hbar } \over {z-w}}\ \ \ \ \ \ \ \
\ \ \ \ \ \ \ \ &\ \ \ \ \ \ \ \  b(z){\pi }_b (w) \sim { {\hbar }\over {z-w}}
\cr c(z) {\bar \partial }b(w) = - {\bar \partial }c(z) &b(w) \sim -{{\hbar
}\over z-w },
\cr}\eqno(2.12)
$$
where we have introduced ${\hbar }$ for later convenience in counting loop
orders.

The action (2.11) is invariant under the following BRST transformations:
$$
\eqalignno{&  \delta {\varphi }^i=c\, {\partial {\varphi }^i} ,  &(2.13a) \cr
& \delta h = {\pi }_b  , &(2.13b)  \cr
& \delta c = c\, \partial c   , &(2.13c) \cr
& \delta {\pi }_c = T_{\rm mat} + T_{\rm gh} , &(2.13d) \cr
& \delta b = {\pi }_h  ,  &(2.13e) \cr
& \delta {\pi }_b = 0  ,  &(2.13f) \cr
& \delta {\pi }_h = 0 ,  &(2.13g) \cr }
$$
where $T_{\rm gh}$ no longer takes the form (2.5b), but is now instead
$$
T_{\rm gh}= -2 {\pi }_c\, \partial {c} - {\partial {\pi }_c }\, c.
\eqno (2.14)
$$
These transformations are all nilpotent at the classical level, {\it i.e.}\ at
the level of Poisson antibrackets, or by taking OPEs with single contractions
between fields.

  Checking the invariance of the gauge-fixed action under the transformations
(2.13) is made much easier by rewriting (2.11) as
$$
I= {1\over \pi} \int d^2z \Big( -{1\over 2}
{\bar\partial{\varphi }^i }\,\partial {\varphi }^i
+ {\pi }_h \,{\bar \partial }h - {\pi }_b \,{\bar \partial }b
-{\pi }_c \,{\bar \partial }c - \delta ( h \,{\pi }_c ) \Big) . \eqno(2.15)
$$
 From the action (2.15), we can see that the Hamiltonian density ${\cal H}=
\delta( h\,{\pi }_c )$ is BRST trivial, and hence is invariant as a consequence
of the nilpotence of the transformations (2.13). The kinetic terms
constituting the remainder of (2.15) are also invariant because terms of the
form $\pi_\phi\, \bar\partial\phi$ are invariant up to a total
$\bar\partial$ derivative under {\it arbitrary} canonical
transformations.\footnote{$^{\ast}$}{\tenfoot Note that in the worldsheet
coordinates $z$, $\bar z$ that we are using, the scalar kinetic term
$\bar\partial\varphi^i\partial\varphi^i$ is already in first-order form with
respect to the ``evolution'' $\bar\partial$ derivatives, so we have not
bothered to introduce conjugate momenta for the $\varphi^i$. Strictly
speaking, in the canonical formalism one should introduce momenta $\pi^i$,
and then should deal with the resulting constraint
$\pi^i=-\ft12\partial\varphi^i$ using the Dirac-bracket formalism. In
practice, we shall find it simpler to do our explicit calculations using
the quantum operator products (2.12), restricted to single contractions
when discussing the classical or tree level. Further details on the
canonical formalism for chiral theories may be found in Ref.\ [14].}

  That the transformations (2.13) are now canonical is one of the main
benefits of our reformulated BRST quantisation procedure. The
generator of these transformations is the BRST charge $Q$. This may be obtained
from (2.13, 2.15) using Noether's theorem, by supplying an anticommuting
transformation parameter $\lambda$ which is allowed to depend on $\bar
z$ and then collecting the factors multiplying $\bar\partial\lambda$ in the
BRST variation of the action, giving
$$
  Q= \int dz \Big( c\,(T_{\rm mat}+{1\over 2}T_{\rm gh}) + {\pi }_h\, {\pi }_b
\Big).
\eqno(2.16)
$$
One may verify, using the OPE relations (2.12), that the desired rule (2.8) is
now correctly obtained for the transformations (2.13a--g) of all fields,
including the gauge field $h$.

  Having a well-defined BRST formalism in hand, we now proceed to
calculate the Virasoro anomaly from the BRST Ward identity.

\bigskip
\noindent{\it Virasoro Anomaly in Non-critical Dimensions}
\medskip

  To set the stage for our discussion of the worldsheet gravity anomalies, let
us first review the BRST treatment of anomalies, following Ref.\ [16].

  Suppose ${\cal L}_{\rm gf} $ is a BRST-invariant gauge-fixed Lagrangian
density. In order to calculate correlation functions and derive the BRST form
of the Ward identity, one has to introduce three kind of sources: $J_{{\phi
}^i}$ and $K_{{\phi }^i}$ for the fields and their variations and $L$ for the
anomaly, the extended Lagrangian density
${\cal L}_{\rm ext,\, anom}$ then being given by
$$\eqalignno{
{\cal L}_{\rm ext}&={\cal L}_{\rm gf} + J_{{\phi }^i} \,{\phi }^i +
K_{{\phi }^i} \,\delta {\phi }^i , &(2.17a)\cr
{\cal L}_{\rm ext,\,anom}&= {\cal L}_{\rm ext } + L {\triangle } , &(2.17b)\cr}
$$
where ${\phi }^i$ generically denotes all fields and ${\triangle}$
denotes the anomaly. The partition function with dependence upon these
sources is then defined as
$$
\eqalign{
{\cal Z}(J_{{\phi }^i},K_{{\phi }^i},L)
=\int {\cal D}{\phi }^i
e^{-\int d^2z {\cal L}_{\rm ext,\,anom}}  .}
\eqno (2.18)
$$
The generating function ${\cal W}$ and the effective action functional
$\Gamma$ are defined respectively as
$$
 \eqalignno{ &{\cal W}_{(J_{{\phi }^i},K_{{\phi }^i},L)} =
{\rm ln}{\cal Z}(J_{{\phi }^i},K_{{\phi }^i},L) , &(2.19a) \cr
&{\Gamma }({\phi }^i,K_{{\phi }^i},L)=
{\cal W}(J_{{\phi }^i},K_{{\phi }^i},L) - \int d^2z J_{{\phi }^i}
{\phi }^i . &(2.19b) \cr }
$$

 The path-integral measure ${\cal D}{\phi }$ is not generally invariant under
Weyl transformations. Consequently, it is also not generally invariant under
our
BRST transformations, since the residual Virasoro symmetry (2.2) that is left
unfixed by our preliminary gauge choice (1.12) is a composite of a Weyl
transformation and compensating diffeomorphisms. In consequence, although the
action $\int {\cal L}_{\rm ext}$ is BRST invariant, the partition function is
generally not and so one has a BRST anomaly:
$$
\eqalign{ {\cal Z}-{\cal Z'}&= \int {\cal D}{\phi }^i J_{\phi^i}
\delta {\phi }^i e^{-\int d^2z {\cal L}_{\rm ext,\ anom} } \cr
&={\cal Z} J_{{\phi }^i} {\delta {\cal W} \over \delta K_{\phi^i}}
=-{\cal Z} {\delta \Gamma \over \delta {\phi }^i}
{\delta \Gamma \over \delta K_{{\phi }^i}} \cr
&=-{\cal Z} {\triangle } \cdot \Gamma =-{\cal Z} {\partial \Gamma \over
\partial L}=-{\partial {\cal Z} \over \partial L}, \cr}
\eqno (2.20)
$$
where we are using De Witt notation, so that repeated indices denote both
Einstein summation and integration over arguments. From the identities (2.20),
we can extract two very useful equations. The first of these is
$$
{\delta \Gamma \over \delta {\phi }^i}
{\delta \Gamma \over \delta K_{{\phi }^i}} ={\triangle } \cdot \Gamma \equiv
{\partial \Gamma \over\partial L},
\eqno (2.21)
$$
where ${\triangle } \cdot \Gamma $ denotes the set of all 1PI diagrams with an
insertion of the composite anomaly operator ${\triangle}$. The second
equation, obtained for $L\rightarrow 0$, is
$$
\int {\cal D}{\phi }^i J_{{\phi }^i} \delta {\phi }^i e^{-\int d^2z
{\cal L}_{\rm ext}} = -\int {\cal D} {\phi }^i {\triangle } e^{-\int d^2z
{\cal L}_{\rm ext}} .
\eqno (2.22)
$$
Equations (2.21) and (2.22) are different expressions of the anomalous Ward
identity.

  Returning to our specific case of chiral Virasoro gravity, the extended
Lagrangian density ${\cal L}_{\rm ext}$ becomes\cramp
$$
\eqalignno{{\cal L}_{\rm ext} = & -{1\over 2}
{\bar\partial{\varphi }^i }\,\partial {\varphi }^i
+ {\pi }_h\, {\bar \partial }h - {\pi }_b \,{\bar \partial }b
-{\pi }_c \,{\bar \partial }c -{\pi }_b\, {\pi }_c
-h \,( T_{\rm mat} + T_{\rm gh}  ) + J_{{\varphi }^i } \,{\varphi }^i +
J_{h} \,h + J_{c} \,c + J_{b}\, b \cr
&\ \  + J_{{\pi }_c}\, {\pi }_c + J_{{\pi }_b}\, {\pi }_b +
K_{{\varphi }^i}\,c \,{\partial } {\varphi }^i + K_{h}\, {\pi }_b +
K_{b}\, {\pi }_h + K_{c} \,c\, \partial c + K_{{\pi}_c}\,
( T_{\rm mat} + T_{\rm gh} ).
&(2.23)\cr }
$$\uncramp
Expanding loopwise in powers of $\hbar$, we have
$$
\Gamma ={\Gamma }_0 + \hbar {\Gamma }_1 + {\hbar }^2 {\Gamma }_2
+\cdots , \ \ \ \ \ \ \  \triangle = \hbar {\triangle }_1 + {\hbar }^2
{\triangle }_2 +\cdots .
\eqno (2.24)
$$
where ${\Gamma }_0 $ is the extended action $I_{\rm ext}$, ${\Gamma }_1$ is the
one loop correction to the effective action, ${\triangle }_1 $ is the
one loop contribution to the anomaly, etc.

  We shall be interested here firstly in the one-loop contributions to the
anomalous Ward identity (2.21), which at order $\hbar$ becomes
$$
{\delta {\Gamma }_0 \over \delta {\phi }^i}\,
{\delta {\Gamma }_1 \over \delta K_{{\phi }^i}}
+{\delta {\Gamma }_0 \over \delta K_{{\phi }^i}}\,
{\delta {\Gamma }_1 \over \delta {\phi }^i} = {\triangle }_1 .
\eqno (2.25)
$$
Explicit evaluation shows that the only non-vanishing contributions to the
anomaly come from
$$
\eqalign{ {\delta {\Gamma }_0 \over \delta {\pi }_c }
{\delta {\Gamma }_1 \over \delta K_{{\pi }_c}}
=&{1\over {\pi }^2\hbar^2} \int d^2z d^2w {\bar \partial }c(z)
  \Big\langle {\Big( {1\over 2} {\partial } {\varphi }^i {\partial } {\varphi
}^i +2 {\pi }_c \,{\partial }c + {\partial } {\pi }_c \,c \Big) }(z) \cr
&\ \ \ \ \ \ \ \ \ \  {\Big(  ( h - K_{{\pi }_c} ) ( {1\over 2}
{\partial } {\varphi }^i {\partial } {\varphi }^i
+2 {\pi }_c \,{\partial }c + {\partial } {\pi }_c \,c ) \Big) }(w)
\Big\rangle + \ldots, \cr }
\eqno (2.26)
$$
where the angle brackets $\langle \ \rangle$ denote the OPE of the bracketed
fields and the omitted terms do not give local contributions. After some
algebra, one gets the anomaly
$$
\eqalign { {\triangle }_1 &= {\delta {\Gamma }_0 \over \delta {\pi }_c }\,
{\delta {\Gamma }_1 \over \delta K_{{\pi }_c}}+\ldots \cr
&=-{1\over 2 {\pi }^2 } ( 26-D ) \int d^2z d^2w
{1 \over (z-w)^4 } {\bar \partial } c(z) \Big( h(w) - K_{\pi_c}(w)
\Big) \cr
&={ 1\over 12 {\pi } } ( 26-D ) \int d^2z d^2w
{ {\partial }_{z}^3 \over {\bar \partial }_{z} } {\delta }_{(z-w)}\,
 {\bar \partial } c(z) \Big( h(w) - K_{\pi_c}(w) \Big) \cr
&=-{ 1\over 12 {\pi } } ( 26-D ) \int d^2z\,
  c\, \Big( {\partial }^3 h - {\partial }^3 K_{\pi_c}
\Big) . \cr }
\eqno (2.27)
$$

  In order to check whether the form of the anomaly $\triangle$ given
in (2.27) is correct, we need to verify that it satisfies the Wess-Zumino
consistency condition [16].  This condition in general is simply
$$
\Big( \Gamma , \big( \Gamma , \Gamma \big) \Big) =0 ,
\eqno (2.28)
$$
and is a consequence of the Jacobi identity for the antibracket
$(\cdot,\cdot)$, which is defined as
$$
\Big( A  , B  \Big) \equiv
 {\delta A \over \delta {\phi }^i}\,
{\delta B \over \delta K_{{\phi }^i}}
+{\delta A \over \delta K_{{\phi }^i}}\,
{\delta B \over \delta {\phi }^i} ,
\eqno (2.29)
$$
for arbitrary functionals $A$ and $B$. Note that the extended classical action
is BRST invariant, as a consequence of the invariance of the gauge-fixed
action (2.15) and the classical nilpotence of the transformations (2.13). In
antibracket notation, this becomes $({\Gamma }_0 , {\Gamma }_0 ) = 0$. Then,
expanding  $\Gamma$ in a series in $\hbar$, the one-loop Wess-Zumino
consistency condition following from (2.28) becomes
$$
\Big( {\Gamma }_0 , {\triangle }_1 \Big) \equiv
 {\delta {\Gamma }_0 \over \delta {\phi }^i}\,
{\delta {\triangle }_1 \over \delta K_{{\phi }^i}}
+{\delta {\Gamma }_0 \over \delta K_{{\phi }^i}}\,
{\delta {\triangle }_1 \over \delta {\phi }^i} =0 .
\eqno (2.30)
$$
Testing our anomaly (2.27) in this formula, we obtain
$$
\eqalign { \Big( {\Gamma }_0 , {\triangle }_1 \Big) &=
 {\delta {\Gamma }_0 \over \delta {\pi }_c }\, {\delta {\triangle }_1 \over
\delta K_{{\pi }_c } } +{\delta {\Gamma }_0 \over \delta K_{c } } \,{\delta
{\triangle }_1 \over \delta c } +{\delta {\Gamma }_0 \over \delta K_{h } }\,
{\delta {\triangle }_1 \over \delta h } \cr
&=-{1\over 12 \pi } (26-D)
\int d^2z \Big[ ({\pi }_b -{\bar \partial }c -c\, \partial h + {\partial
c}\, h + c\,
\partial K_{{\pi }_c} - \partial c\,
K_{{\pi }_c} ) {\partial }^3 c \cr & \ \ \ \
\ \ \ \ \ \ \ \ \ \ \ \ \ \ \ \ \ \ \ \ \ \ \ \ \ \ \ \ \ \ \ + c\,\partial c
\,({\partial }^3 h - {\partial }^3 K_{{\pi }_c} ) - {\pi }_b \,
{\partial }^3 c \Big]\cr &=0 ,\cr }
\eqno (2.31)
$$
so we verify that the consistency condition indeed is satisfied.
\np
\noindent{\bf 3.  Conventional BRST quantisation of $W_3$ gravity}
\bigskip

  The $W_3$ algebra, originally found by Zamolodchikov [17], in the
conventions of Ref.\ [15] becomes
$$
 \eqalignno{ {\hbar }^{-1} T_{\rm mat}(z)T_{\rm mat}(w) &\sim {
{\partial T_{\rm mat} }\over {z-w}} +
  {{2T_{\rm mat}}\over {(z-w)^2}}
+{\hbar }{{ 1\over 2 } C_{\rm mat}\over {(z-w)^4}} , &(3.1a) \cr
 {\hbar }^{-1}T_{\rm mat}(z)W_{\rm mat}(w) &\sim {{\partial W_{\rm mat}
}\over {z-w}} +
  { {3W_{\rm mat}}\over {(z-w)^2} } , &(3.1b) \cr
 {\hbar }^{-1}W_{\rm mat}(z)W_{\rm mat}(w) &\sim {1\over {z-w}} \Big( {1\over
{15}} {\hbar } {\partial }^3 T_{\rm mat} + a {\partial \Lambda  } \Big )
\cr &+ {1\over {(z-w)^2}} \Big( {3\over {10}} {\hbar } {\partial }^2 T_{\rm
mat} + 2 a \Lambda  \Big ) \cr &+ {\hbar }{{\partial T_{\rm mat}
}\over {(z-w)^3}} +
 {\hbar } {{2T_{\rm mat}}\over {(z-w)^4}} + {\hbar }^2 {{{1\over 3} C_{\rm
mat} }\over {(z-w)^6}} , &(3.1c) \cr }
$$
where $a={16\over {22+5C_{\rm mat}}}$ and  $\Lambda $ is a composite current
given by the normal-ordered product\footnote{$^{\dag}$}{\tenfoot Normal
ordering is denoted here by round brackets $(\ )$.}
$$
  \Lambda = (T_{\rm mat}T_{\rm mat})-{3\over 10} {\partial }^2 T_{\rm mat}.
\eqno(3.2)
$$
The matter currents $T_{\rm mat}$ and $W_{\rm mat}$ represent respectively the
spin 2 and spin 3 generators of the $W_3$ algebra; their explicit realisations
in terms of $n$ scalar fields ${\varphi }^i$ are
 $$
\eqalignno{ T_{\rm mat}&=
-{1\over 2} \partial {\varphi }^i \,\partial {\varphi }^i -
\sqrt{\hbar }  {\alpha }_i\, {\partial }^2 {\varphi }^i ,  &(3.3a) \cr
  W_{\rm mat}&= -{1\over 3} d_{ijk} \,\partial {\varphi }^i \,\partial
{\varphi}^j
   \,\partial {\varphi }^k - \sqrt{\hbar}\, e_{ij} \,\partial {\varphi }^i
   \,{\partial }^2 {\varphi }^j - {\hbar}\, f_i\,{\partial }^3 {\varphi }^i,
 &(3.3b) \cr }
$$
where the ${\varphi }^i$ satisfy the OPE
$$
{\partial }{\varphi }^i(z){\partial }{\varphi }^j(w) \sim
{-{\hbar }{\delta }^{ij} \over (z-w)^2 }.
\eqno(3.4)
$$
In order to obtain a realisation of the $W_3$ algebra\footnote{$^{\ddag}$}
{\tenfoot Here, unlike in the Virasoro case, we
must include background charges in order to have a multi-field realisation
[18]; without background charges, one has only the original (non-critical)
two-field realisation of Ref.\ [17].} (3.1), the constants
${\alpha }_i, d_{ijk}, e_{ij}$ and
$f_i$  must satisfy the following relations found in Ref.\ [18]:
$$
\eqalignno{&  d_{ijj} - 6 e_{ij}\, {\alpha }_j + 6 f_i = 0 , &(3.5a) \cr
& e_{(ij)} - d_{ijk}\, {\alpha }_k = 0 , &(3.5b) \cr
& 3 f_i - {\alpha }_j\, e_{ji} = 0 , &(3.5c) \cr
& d_{ikl}\,d_{jkl} + 6 d_{ijk}\, f_k -3 e_{ik}\, e_{jk}
 = {1\over 2} {\delta }_{ij} , &(3.5d) \cr
&  d_{(ij}{}^m\, d_{kl)m} =
{1\over 2} a \,{\delta }_{(ij} {\delta }_{kl)} , &(3.5e) \cr
& d_{ijk} ( e_{lk} - e_{kl} ) + 2 e_{(i}{}^l\, d_{j)kl} = a\, {\alpha }_k
\,{\delta }_{ij} . &(3.5f) \cr }
$$
Two useful consequences of this set of equations which will
be useful in later calculations are
$$
\eqalignno {& e_{ii} + 12 {\alpha }_i\, f_i = 0 , &(3.5g) \cr
& C_{\rm mat}=-2d_{ijk}\,d_{ijk}-18e_{ij}\,e_{ij}-12e_{ij}\,e_{ji}
-360f_i^2 . &(3.5h) \cr }
$$
The central charge $C_{\rm mat} $ is given for the realisation (3.3) by
$$
C_{\rm mat} = n + 12 {\alpha }_i {\alpha }_i.
\eqno (3.6)
$$

  The BRST charge for this system was given by J.\ Thierry-Mieg [19] as
follows:
$$
  Q= \int dz \Big( c(T_{\rm mat}+{1\over 2}T_{\rm gh})+ \gamma (W_{\rm mat}+
{1\over 2}W_{\rm gh}) \Big),
\eqno(3.7)
$$
in which the ghost currents $T_{\rm gh}$ and $W_{\rm gh}$ are given by
$$
\eqalignno{ T_{\rm gh}&= -2 b \,\partial {c}
- {\partial {b} }\, c -3 \beta\, {\partial \gamma }
         -2 {\partial \beta } \,\gamma , &(3.8a) \cr
W_{\rm gh}&= -{\partial \beta } \,c - 3 \beta \,{\partial c} -
         a [{\partial (b\, \gamma \,T_{\rm mat})} + b\, {\partial \gamma }\,
         T_{\rm mat} ]\cr
         & + { (1-17 a)\over 30} {\hbar }(2 \gamma \,{\partial }^3 b +
           9 {\partial \gamma } \,{\partial }^2 b +
           15 {\partial }^2 \gamma \,\partial b +
           10 {{\partial }^3 \gamma } \,b ). &(3.8b) \cr}
$$
The ghost-antighost pairs (c,b) and ($\gamma $, $\beta $) correspond
respectively to the $T$ and $W$ generators. They satisfy the following
OPEs
$$
c(z)b(w) \sim {{\hbar } \over {z-w}}\ ; \ \ \ \ \ \ \ \ \ \ \ \ \ \ \ \
\gamma (z) \beta (w) \sim { {\hbar }\over {z-w}}.
\eqno(3.9)
$$
Note that here, unlike in the Virasoro case where the ghost current $T_{\rm
gh}$
forms a separate realisation of the Virasoro algebra, the ghost currents (3.8)
do not form a separate realisation of the $W_3$ algebra.

  The BRST charge (3.7) corresponds to the conventional BRST action for $W_3$
gravity [15]\cramp
$$
\eqalignno{ I&={1\over \pi }\int d^2z {\cal L}\cr
&={1\over \pi } \int d^2z \Big( -{1\over 2}
{\bar\partial{\varphi }^i }\,\partial {\varphi }^i
- h\, T_{\rm mat} - B\, W_{\rm mat} + \delta [ b \,( h - h_{back} ) +
\beta\,( B - B_{back} ) ] \Big) &(3.10a)\cr
&={1\over \pi } \int d^2z
\Big( -{1\over 2}
{\bar\partial{\varphi }^i }\,\partial {\varphi }^i
- b \,{\bar\partial {c} }
- \beta \,{\bar\partial\gamma } + {\pi }_h \,(h-h_{back})
+ {\pi }_B \,(B-B_{back}) \cr
&\phantom{={1\over \pi } \int d^2z}\ \ \ \ \  - h\,(T_{\rm mat}+T_{\rm gh}) -
B\,(W_{\rm mat}+W_{\rm gh}) \Big) , &(3.10b)\cr}
$$\uncramp
where the conventional gauge conditions for the spin-2 and spin-3 gauge fields
$h$ and
$B$ are $h=h_{back}$ and $B=B_{back}$, imposed by the Lagrange multipliers
${\pi }_h$ and ${\pi}_B$. The transformation rules of the fields in (3.10) are
given by\cramp
$$
\eqalignno{ \delta {\varphi }^i =&c\, {\partial {\varphi }^i} +
d_{ijk} \,\gamma\,
{\partial {\varphi }^j} {\partial {\varphi }^k} + a\, b \,\gamma\,
{\partial \gamma }\, {\partial {\varphi }^i} \cr
&+{\sqrt \hbar } \Big( -{\alpha }_i \,\partial c + (e_{ij}-e_{ji}) \gamma\,
{\partial }^2 {\varphi }^j - e_{ji} \,\partial \gamma \,\partial {\varphi }^j
-a \,{\alpha }_i \,\partial (b\, \gamma \,\partial \gamma ) \Big) +
\hbar\, f_i\, {\partial }^2 \gamma  , &(3.11a) \cr
 \delta h =&{\bar \partial }c + c\, \partial h - {\partial c}\, h
- {a \over 2} ( \gamma \,\partial B - {\partial \gamma } \,B )
\partial {\varphi }^i \partial {\varphi }^i - a {\sqrt \hbar } ( \gamma\,
\partial B - {\partial \gamma } \,B ) {\alpha }_i \,{\partial }^2
{\varphi }^i \cr
& + {{1-17 a}\over 30 } \hbar ( 2 \gamma \,{\partial }^3 B - 3 \partial
\gamma\,
{\partial }^2 B + 3 {\partial }^2 \gamma \,\partial B - 2 {\partial }^3 \gamma
\,B) , &(3.11b) \cr
 \delta B =&{\bar \partial }\gamma +c\, \partial B - 2 {\partial
c}\, B + 2 \gamma \,\partial h - \partial \gamma\, h ,  &(3.11c) \cr
 \delta c =& c\,
\partial c -{a\over 2}\, \gamma \,\partial \gamma\,
\partial {\varphi }^i \partial {\varphi }^i - a\, {\sqrt \hbar }\, {\alpha }_i
\,\gamma\, \partial \gamma \,{\partial }^2 {\varphi }^i +{{1-17 a}\over 30}
\hbar ( 2
\gamma \,{\partial }^3 \gamma - 3 \partial \gamma \,{\partial }^2 \gamma ) ,
&(3.11d) \cr
 \delta \gamma =& c\, \partial \gamma - 2 \partial c\, \gamma ,
&(3.11e) \cr
 \delta b = & {\pi }_h ,  \ \ \ \ \ \ \ \  \delta \beta = {\pi }_B ,
\ \ \ \ \ \ \ \
\delta {\pi }_h = 0 , \ \ \ \ \ \ \ \  \delta {\pi }_{B } = 0 . &(3.11f,g,h,i)
\cr }
$$\uncramp

  In saying that (3.10) corresponds to (3.7), one needs to be careful about
the logical status of this correspondence. The currents $T_{\rm mat}$, $W_{\rm
mat}$, $T_{\rm gh}$ and $W_{\rm gh}$ certainly appear to be the natural
elements
to be extracted from the structure of the full quantum BRST charge (3.7) for
constructing the renormalized action (3.10), once one has realised that a
consistent Lagrangian quantisation must be based upon the original quantum
$W_3$
algebra (3.1) instead of some deformation of this algebra. Indeed, the work of
Ref.\ [15] started from this natural guess. But the relation between (3.10)
and (3.7) is not algorithmic. The results of Ref.\ [15] may be summarised as
the demonstration, by explicit Feynman-diagram evaluation of the
low-order anomalies, that for values of the background charges corresponding to
a realisation of (3.1) with $C_{\rm mat}=100$ the guess (3.10) actually works,
{\it i.e.}\ that all the matter-dependent and universal anomalies do cancel.
The renormalized action (3.10) is itself invariant under the transformations
(3.11) only up to order $\hbar^{\sfrac12}$. It fails to be invariant at order
$\hbar$ precisely as needed so as to cancel the local anomalous terms arising
in the BRST Ward identity from variations of {\it non-local} contributions to
the effective action $\Gamma$. The invariance of (3.10) under (3.11) at orders
$\hbar^0$ and $\hbar^{\sfrac12}$ allows one to investigate a {\it
partially}-anomalous generalisation of the results of [15] with $C_{\rm mat}\ne
100$. In this partially-anomalous case, the status of the central charge
$C_{\rm mat}$ as an independent free parameter related to the order
$\hbar^{\sfrac12}$ background charges $\alpha^i$ {\it via} (3.5, 3.6) makes it
possible to abandon the requirement of cancellation of the universal anomalies
(depending only on $h$ and $B$ and the ghosts and antighosts) while still
insisting nonetheless upon cancellation of all anomalies depending on the
matter fields $\varphi^i$. Implementing this noncritical scheme is possible but
very cumbersome in the theory defined by the conventional BRST action (3.10)
[13], but the implementation will become much more transparent in our
reformulated BRST quantisation procedure.

  It is clear upon inspection of the BRST charge (3.7) and the transformation
rules (3.11) that the transformation rules for the gauge fields $h$ and
$B$ cannot be obtained directly from the BRST charge (3.7). This is as in the
Virasoro case, and one can only obtain the transformations of $h$ and $B$ by
requiring the invariance of the theory. This is what was done in the critical
$C_{\rm mat}=100$ theory [15], yielding the result that the gauge fields
must transform in the coadjoint representation of the algebra. Clearly,
however, the choice of transformations for $h$ and $B$ is trickier in the
noncritical $C_{\rm mat}\ne 100$ theory, since one obviously cannot require
invariance of the theory in this case, and the renormalisation corrections to
the $h$ and $B$ transformations must be determined by requiring the absence of
matter-dependent anomalies in the anomalous Ward identity. This requirement
presumably leads once again to the requirement that the gauge fields transform
in the coadjoint representation as in (3.11$b$,$c$), but we are not aware of a
direct verification of this fact in the conventional formulation (3.10) of the
noncritical theory. Once again, this issue will become greatly simplified in
our
reformulated quantisation procedure, where $h$ and $B$ will be treated in the
same way as all other fields.

  Before leaving the conventional BRST formulation of $W_3$ gravity, we recall
[11] that the BRST transformations (3.11) have a structure that is not directly
obtained by the standard simple prescription of replacing the parameters of
the classical $W_3$ gauge transformations by ghosts. The classical chiral
$W_3$ gravity action is [20]
$$
  I_{\rm class}={1\over \pi } \int d^2z \Big(-{1\over 2}
{\bar\partial{\varphi }^i }\,\partial {\varphi }^i + {1\over 2} h\,
\partial {\varphi }^i \partial {\varphi }^i + {1\over 3}
 B\, d_{ijk} \,\partial {\varphi }^i \partial {\varphi }^j
\partial {\varphi }^k \Big) ,
\eqno (3.12)
$$
and it is invariant under the following infinitesimal transformations
$$
\eqalignno{&  \delta {\varphi }^i =  {\varepsilon }\,
{\partial {\varphi }^i} + d_{ijk}\, {\eta }\, \partial {\varphi }^j
\partial {\varphi }^k ,  &(3.13a) \cr
& \delta h ={\bar \partial }{\varepsilon } + {\varepsilon }\, \partial h
- {\partial {\varepsilon }}\, h
- {a \over 2} ( \eta\, \partial B - {\partial \eta } \,B )
\partial {\varphi }^i \partial {\varphi }^i  ,  &(3.14b) \cr
& \delta B ={\bar \partial }\eta +{\varepsilon } \,\partial B -
2 {\partial {\varepsilon }}\,B
+ 2 \eta\, \partial h - \partial \eta \,h , &(3.14c) \cr }
$$
where $\varepsilon$ and $\eta$ are infinitesimal parameters for the Virasoro
and
spin-3 transformations, respectively. One would normally expect the
corresponding BRST transformations to be obtained simply by replacing
$\varepsilon$ by the spin-2 ghost $c$ and $\eta$ by the spin-3 ghost
$\gamma$. However, comparison with the transformation (3.10a) of $\varphi^i$
shows that the actual transformation contains an unexpected term
$b\,\gamma\,\partial\gamma\,\partial\varphi^i$. As a consequence, the classical
action (3.12) is not invariant under the BRST transformations
(3.11).\footnote{$^{\ast}$}{\tenfoot It is, of course, still true that the
$\hbar$-independent terms in the complete action (3.10$b$) are invariant
under the $\hbar$-independent terms in the BRST transformation rules (3.11).
The slightly unusual new feature in a case such as $W_3$ gravity, where
the algebra is non-linear, is that the $\hbar$-independent terms in the
ghost kinetic terms and gauge-fixing terms in the total action play a r\^ole
in ensuring invariance under BRST transformations at the classical level.
One could consider a rather trivial linear classical algebra, in which the
Poisson bracket of $W_{\rm mat}$ with $W_{\rm mat}$ were zero, rather than
being proportional to $(T_{\rm mat})^2$.  This would correspond to setting the
constant $a$ to zero in the classical action and BRST transformation rules,
under which circumstances the unusual terms that do not come from a simple
replacement of parameters by ghosts would disappear.}

  The non-standard term in the $\delta\varphi^i$ transformation is related to
another peculiarity of the BRST transformations (3.11). Taking the
$\hbar\rightarrow 0$ limit of (3.11) and calculating
$\delta^2$ on the various fields, one finds that
the transformations fail to be nilpotent on $h$ and $\varphi^i$. For example,
one finds after some algebra that
$$
\eqalign{ {\delta }^2 h = a \,\gamma \,\partial \gamma \,\partial {\varphi }^i
\Big[ & -{\bar \partial }\partial {\varphi }^i + h \,{\partial }^2
{\varphi }^i + \partial h\, \partial {\varphi }^i \cr
 & \ \ + d_{ijk} \,\partial B\,
\partial {\varphi }^j \partial {\varphi }^k + d_{ijk} \,B \,\partial (
\partial {\varphi }^j \partial {\varphi }^k ) + a\, {\partial }^2
\gamma\, b\, B \,\partial {\varphi }^i \Big]. \cr}
\eqno (3.15)
$$
Similarly, in $\delta^2\varphi^i$ one has, amongst other terms, a term
$a\,\pi_h\,\gamma\,\partial\gamma\,\partial\varphi^i$ that is nowhere
canceled. Both of these non-closure expressions, however, vanish upon use of
the classical field equations. For example, the equation of motion for
$\varphi^i$ is
$$
\eqalign{ {\partial {\cal L} \over \partial {\varphi }^i } -
\partial {\partial {\cal L} \over \partial \partial {\varphi }^i } =&
{\bar \partial }\partial {\varphi }^i - h\, {\partial }^2
{\varphi }^i - \partial h \,\partial {\varphi }^i
  - d_{ijk} \,\partial B\,
 \partial {\varphi }^j \partial {\varphi }^k \cr
&\ \  - d_{ijk}\, B \,\partial (
\partial {\varphi }^j \partial {\varphi }^k ) + a \,\partial (
B\, b\, \partial \gamma \,\partial {\varphi }^i )=0, \cr }
\eqno (3.16)
$$
which causes the terms in the square brackets in (3.15) to vanish. Similarly,
the equation of motion for the gauge field $h$ is
$$
\pi_h-(T_{\rm mat}+T_{\rm gh})=0,\eqno(3.17)
$$
which causes the non-closure terms in $\delta^2\varphi^i$ to vanish.
Nonetheless, the transformations (3.11) do express an invariance of the
gauge-fixed action (3.10) because the terms arising from the
variation of $h$ in $\delta(b\,h)$ in (3.10$a$) due to the $\delta^2h$
off-shell
non-closure are canceled by the extra term
$a\,b\,\gamma\,\partial\gamma\,\partial\varphi^i$ in the transformation of
$\varphi^i$. The off-shell non-closure of the BRST algebra and the
corresponding complication of the BRST transformations is strongly reminiscent
of the BRST formulation of supergravity theories prior to the introduction of
auxiliary fields. As in that case [21], curing
these problems will require us to find a new formalism that achieves full
off-shell closure of the BRST algebra.

\bigskip
\noindent{\bf 4. Canonical BRST quantisation of $W_3$ gravity}
\bigskip

  In this section, we shall quantise chiral $W_3$ gravity using our
reformulated
BRST construction based upon derivative gauge conditions for the spin-2 and
spin-3 gauge fields. We shall calculate all the anomalies at order
$\hbar$ together with the local anomaly at order $\hbar^2$, which contains the
spin-3 anomaly. We shall show that these anomalies satisfy the Wess-Zumino
consistency condition as required. In this discussion, we shall present the
BRST formulation for a general model based upon a realisation of the $W_3$
algebra with an arbitrary value of the matter central charge $C_{\rm mat}$,
adjustable by an appropriate choice of the background-charge terms. Of course,
from the point of view of noncritical $W_3$ gravity, the most natural choice
for the values of these central charges might be considered to be zero, which
is also covered by the general discussion that we shall give.

  The gauge conditions that we shall choose for chiral $W_3$ gravity are
$\bar\partial h = 0$ and $\bar\partial B = 0$, naturally generalising our
discussion in the Virasoro-gravity case. This BRST quantisation procedure in
the $W_3$ case proceeds now along lines strictly parallel to our Virasoro
discussion. Accordingly, we now shorten our $W_3$ discussion by presenting
directly the fully-renormalized gauge-fixed chiral $W_3$-gravity action,
including background-charge terms with parameters $\alpha^i$ chosen so as to be
consistent with (3.5) but in general corresponding to $C_{\rm mat}\ne
100$:\footnote{$^{\dag}$}{\tenfoot In the case of the minimal two-scalar
realisation of the $W_3$ algebra with $\varphi^1$ playing the r\^ole of the
Virasoro Liouville field and $\varphi^2$ playing the analogous r\^ole for the
spin-3 symmetry, equations (3.5) require $\alpha_1=\sqrt3\alpha_2$ in order to
obtain a realisation of the $W_3$ algebra. This relation leaves undetermined
one
background-charge parameter, corresponding to the unfixed value of the
matter-sector central charge $C_{\rm mat}$.}
$$
\eqalign{I =&{1\over \pi} \int d^2z {\cal L} \cr
=& {1\over \pi} \int d^2z \Big( -{1\over 2}
{\bar\partial{\varphi }^i }\,\partial {\varphi }^i
- h\, T_{\rm mat} - B\, W_{\rm mat} + {\pi }_h\, {\bar \partial }h \cr
&- b\, {\bar \partial }[{\bar \partial }c +
c\, \partial h - {\partial c}\, h
- {a \over 2} ( \gamma \,\partial B - {\partial \gamma } \,B )
\partial {\varphi }^i \partial {\varphi }^i
- a {\sqrt \hbar } ( \gamma\, \partial B - {\partial \gamma }\, B )
{\alpha }_i \,{\partial }^2 {\varphi }^i \cr
&\ \ \ \ \ \ \ \ \ \  + {{1-17 a}\over 30 } \hbar
( 2 \gamma\, {\partial }^3 B - 3 \partial \gamma\,
{\partial }^2 B + 3 {\partial }^2 \gamma\, \partial B - 2 {\partial }^3 \gamma
\, B ) ] \cr
&\ \ \ \ \ \ \ \ \ \  - \beta\, {\bar \partial } [{\bar \partial }\gamma +c
\,\partial B - 2 {\partial c}\,B
+ 2 \gamma \,\partial h - \partial \gamma \,h ]
 \Big), \cr}\eqno(4.1)
$$
where $T_{\rm mat}$ and $W_{\rm mat}$ are given in (3.3). As in the Virasoro
case, we now reduce this action to first-order form by introducing momenta
conjugate to $c,\ b,\ \gamma,\ \beta$:
$$
\eqalignno{{\pi }_c &={{\partial }{\cal L}\over \partial {\bar \partial }c}=
-{\bar \partial }b \cr
{\pi }_b &={{\partial }{\cal L}\over \partial {\bar \partial }b}=
{\bar \partial }c +
c \,\partial h - {\partial c}\, h
- {a \over 2} ( \gamma \,\partial B - {\partial \gamma } \,B )
\partial {\varphi }^i \partial {\varphi }^i  \cr
&- a {\sqrt \hbar } ( \gamma \,\partial B - {\partial \gamma }\, B )
{\alpha }_i \,{\partial }^2 {\varphi }^i
+{{1-17 a}\over 30 } \hbar
( 2 \gamma\, {\partial }^3 B - 3 \partial \gamma\,
{\partial }^2 B + 3 {\partial }^2 \gamma\, \partial B - 2 {\partial }^3 \gamma
\,B )  \cr
{\pi}_{\gamma} &={\partial {\cal L}\over \partial {\bar \partial} \gamma }
= - {\bar \partial } \beta  \cr
{\pi }_{\beta} &={{\partial }{\cal L}\over \partial {\bar \partial }\beta }
={\bar \partial }\gamma +c \,\partial B - 2 {\partial c}\,B
+ 2 \gamma\, \partial h - \partial \gamma \,h . &(4.2)\cr}
$$
Using these definitions, the gauge-fixed action (4.1) may be put into
first-order form (once again considering the off-diagonal kinetic term
${\bar\partial{\varphi }^i }\,\partial {\varphi }^i$ to be of first order in
$\bar\partial$ derivatives):
$$
\eqalign{I = {1\over \pi} \int d^2z \Big( & -{1\over 2}
{\bar\partial{\varphi }^i }\,\partial {\varphi }^i
+ {\pi }_h \,{\bar \partial }h + {\pi}_B \,{\bar \partial }B
- {\pi }_b \,{\bar \partial }b
-{\pi }_c \,{\bar \partial }c -{\pi}_{\beta}\, {\bar \partial }\beta
-{\pi}_{\gamma} \,{\bar \partial }\gamma \cr
&\ \  -{\pi }_b \,{\pi }_c
-{\pi}_{\beta}\, {\pi}_{\gamma}    -h \,( T_{\rm mat} + T_{\rm gh} )
-B \,( W_{\rm mat}+W_{\rm gh} ) \Big), \cr }\eqno(4.3)
$$
where $T_{\rm gh}$ and $W_{\rm gh}$ are no longer given by (3.8$a$,$b$) but
now take forms involving the conjugate momenta,
$$
\eqalignno{ T_{\rm gh}&= -2 {\pi}_c \,\partial {c}
- {\partial {\pi}_c }\,c -3 {\pi}_{\gamma} \,{\partial \gamma }
         -2 {\partial {\pi}_{\gamma} }\, \gamma , &(4.4a) \cr
W_{\rm gh}&= -{\partial {\pi}_{\gamma} }\, c - 3 {\pi}_{\gamma}\, {\partial c}
-
         a [{\partial ({\pi}_c\, \gamma \,T_{\rm mat})} +
{\pi}_c \,{\partial \gamma } \,T_{\rm mat} ]\cr
         & + { (1-17 a)\over 30} {\hbar }(2 \gamma\, {\partial }^3 {\pi}_c +
           9 {\partial \gamma } \,{\partial }^2 {\pi}_c +
           15 {\partial }^2 \gamma \,\partial {\pi}_c +
           10 {{\partial }^3 \gamma }\, {\pi}_c ). &(4.4b) \cr}
$$
 From the path-integral generating functional derived from (4.3) we obtain
the following OPE relations:
$$
\tabskip=0pt plus1fil \halign to\displaywidth{
 #&$\displaystyle\hfil#\hfil$&#&$\displaystyle\hfil#\hfil$&#&#\cr
\hfill&\partial {\varphi }^i (z) \partial {\varphi }^i (w) \sim {-
{\hbar }\over (z-w)^2};&\hfill&{\pi }_h (z) h (w) \sim
{{\hbar }\over z-w }\cr
\hfill&c(z){\pi }_c (w) \sim {{\hbar } \over {z-w}} ;&\hfill&b(z){\pi }_b (w)
\sim { {\hbar }\over
{z-w}}\cr
\hfill&{\pi }_B (z) B (w) \sim {{\hbar }\over z-w } ;&\hfill&\gamma
(z){\pi }_{\gamma } (w) \sim {{\hbar } \over {z-w}}&\hfill\cr
\hfill&\beta (z){\pi
}_{\beta } (w) \sim { {\hbar }\over {z-w}};\cr
\hfill&c(z) {\bar
\partial }b(w) = - {\bar \partial }c(z) b(w) \sim -{{\hbar }\over z-w
};&\hfill&\gamma (z) {\bar \partial } \beta (w) = - {\bar \partial
}\gamma (z)\beta (w) \sim -{{\hbar }\over
z-w} . &\hfill&\hfill\rlap{(4.5)}\ \ \ \cr}
$$
The BRST transformations corresponding to the action (4.3) are
$$
\eqalignno{\delta {\varphi }^i &=c\, {\partial {\varphi }^i} + d_{ijk}\,
\gamma\,
{\partial {\varphi }^j} {\partial {\varphi }^k} + a \,{\pi}_c\, \gamma
\,{\partial
\gamma }\, {\partial {\varphi }^i} \cr
&+{\sqrt \hbar } \Big( -{\alpha }_i\,
\partial c + (e_{ij}-e_{ji}) \,\gamma\, {\partial }^2 {\varphi }^j -
e_{ji}\,
\partial \gamma \,\partial {\varphi }^j -a\, {\alpha }_i\,
\partial ({\pi }_c\, \gamma\,
\partial \gamma ) \Big) +
\hbar\, f_i \,{\partial }^2 \gamma ,  &(4.6a) \cr
\delta h &= {\pi}_b  , \qquad
\delta B = {\pi}_{\beta} ,   &(4.6b,c) \cr
\delta c &= c \,\partial c -{a\over 2} \gamma \,\partial \gamma\,
\partial {\varphi }^i \partial {\varphi }^i - a {\sqrt \hbar } \,{\alpha
}_i\,
\gamma \,\partial \gamma \,{\partial }^2 {\varphi }^i +{{1-17 a}\over 30}
\hbar ( 2\gamma \,{\partial }^3 \gamma - 3 \partial \gamma\,
{\partial }^2 \gamma ) ,  &(4.6d) \cr
\delta \gamma &= c \,\partial \gamma - 2 \partial c \,\gamma , &(4.6e) \cr
\delta b &= {\pi }_h  ,\ \ \ \  \delta \beta = {\pi }_B  ,  \ \ \ \
\delta {\pi}_c = T_{\rm mat} + T_{\rm gh}  ,\ \ \ \  \delta {\pi}_{\gamma}=
W_{\rm mat} + W_{\rm gh} , &(4.6f,g,h,i) \cr
\delta {\pi }_h &= 0 ,
\ \ \ \  \delta {\pi}_B = 0 ,\ \ \ \  \delta {\pi}_b = 0 ,
\ \ \ \  \delta {\pi }_{\beta } = 0 ,
&(4.6j,k,l,m) \cr }
$$
where $d_{ijk},\ \alpha_i,\ e_{ij},\ f_i$ are chosen so as to satisfy (3.5).

  Taking the limit $\hbar\rightarrow 0$ in (4.2, 4.3, 4.4, 4.6) and replacing
(4.5) by the corresponding classical Poisson bracket/antibracket relations,
one obtains a classical BRST system that repairs all of the
deficiencies of classical $W_3$ gravity as outlined in section three.
In particular, the classical BRST transformations (4.6) obtained in the limit
$\hbar\rightarrow 0$ are now fully nilpotent without use of classical equations
of motion. This may be verified directly by evaluating $\delta^2$ on the
various fields using (4.6); the checks for ${\pi}_c$ and ${\pi}_{\gamma}$ are
the most algebraically involved.

  The action (4.3) and transformation rules (4.6) that we have given are
fully renormalized and include arbitrary background charges consistent
with (3.5). In considering this general case, one needs to know, in addition to
the invariance of (4.3) at order $\hbar^0$, that the variation of (4.3) under
(4.6) also vanishes at order $\hbar^{\sfrac12}$, and correspondingly that the
nilpotence of the transformations (4.6) is obtained also at order
$\hbar^{\sfrac12}$. These low-order observations will provide the basis of the
Wess-Zumino consistency condition that we shall discuss shortly.

  As we found in (3.10$a$) for the Virasoro case, the action (4.3) may be
rewritten using the BRST transformations (4.6) in ``canonical BRST'' form
$\pi^a\,\bar\partial q^a -\delta\Psi$ as\cramp
$$
\eqalign{I =  {1\over \pi} \int d^2z \Big( -{1\over 2}
{\bar\partial{\varphi }^i }\,\partial {\varphi }^i
+ {\pi }_h \,{\bar \partial }h +{\pi}_B \,{\bar \partial }B
- {\pi }_b \,{\bar \partial }b
-{\pi }_c\, {\bar \partial }c -{\pi}_{\beta} \,{\bar \partial } \beta
-{\pi}_{\gamma} \,{\bar \partial }\gamma
- \delta ( h \,{\pi }_c + B \,{\pi}_{\gamma} ) \Big) , \cr }\eqno(4.7)
$$
\uncramp
identifying the ``gauge fermion'' for our formulation as
$$
\Psi = h\,\pi_c+B\,\pi_\gamma . \eqno(4.8)
$$

  The form (4.7) of the $W_3$ gravity action makes the invariance of the
gauge-fixed action at orders $\hbar^0$ and $\hbar^{\sfrac12}$ manifest, since
the BRST transformations (4.6) now constitute a canonical transformation, and
all the ``kinetic'' $\pi^a\,\bar\partial q^a$ terms in (4.7) (including the
${\bar\partial{\varphi }^i }\,\partial {\varphi }^i$ kinetic term) are
invariant under arbitrary canonical transformations, while the Hamiltonian
density ${\cal H} = \delta\Psi$ is BRST trivial and thus is invariant under
(4.6) up to the order to which (4.6) is nilpotent, {\it i.e.}\ at orders
$\hbar^0$ and $\hbar^{\sfrac12}$.

  The canonical generator of the transformations (4.6) is the BRST charge $Q$
of our reformulated quantisation procedure:
$$
\eqalignno
{Q  = \int dz \Big( & c\, (
T_{\rm mat}+{1\over 2}T_{\rm gh} ) +
\gamma \,( W_{\rm mat} + {1\over 2} W_{\rm gh} ) +
{\pi }_h \,{\pi }_b + {\pi }_B\,
{\pi}_{\beta}    \Big)  &(4.9a) \cr
 = \int dz \Big( & c \,T_{\rm mat} + \gamma\, W_{\rm mat} + {\pi}_h\,
{\pi}_b + {\pi}_B\, {\pi}_{\beta} + {\pi}_{\gamma}\, ( c\, \partial \gamma - 2
\partial c\, \gamma ) \cr
& + {\pi}_c\, ( c\, \partial c + a\, \gamma\, \partial \gamma\,
T_{\rm mat} +{1\over 30} (1-17a) \hbar ( 2 \gamma\, {\partial }^3 \gamma - 3
\partial\gamma\, {\partial }^2 \gamma ) \Big) . &(4.9b) \cr }
$$
This BRST charge differs from that of [12] not only in that it contains the
momenta of our first-order formulation, but also in containing the final two
terms of (4.9$a$), which are new. These new terms generate the transformations
of the spin-2 and spin-3 gauge fields, so (4.9) is now fully canonical.

\bigskip
\noindent{\it The $W_3$ anomaly}
\bigskip

  We formulate the BRST Ward identity of $W_3$ gravity as in section two by
introducing sources $K_{\phi^i}$ for the nonvanishing variations (4.6).  Here
we
denote all fields ${\phi}^i$ and the sources of their variations $K_{{\phi}^i}$
as
$$
\eqalignno{ {\phi}^i &= ( {\varphi}^i,h,{\pi}_h,B,{\pi}_B,
b,{\pi}_b,c,{\pi}_c,{\beta},
{\pi}_{\beta},\gamma, {\pi}_{\gamma} ) , &(4.10a) \cr
K_{{\phi}^i} &=( K_{{\varphi}^i}, K_h, K_B, K_b, K_{\beta},
K_{\gamma}, K_c, K_{{\pi}_c}, K_{{\pi}_\gamma}). &(4.10b) \cr }
$$
The extended classical Lagrangian is the sum of (4.3) plus the source
terms,
$$
\eqalignno{ {\cal L}_{\rm ext} =& {\cal L}_{\rm gf} +
K_{{\phi}^i} \delta {\phi}^i\cr
=& -{1\over 2}
{\bar\partial{\varphi }^i }\,\partial {\varphi }^i
+ {\pi }_h \,{\bar \partial }h + {\pi}_B \,{\bar \partial }B
- {\pi }_b \,{\bar \partial }b
-{\pi }_c \,{\bar \partial }c -{\pi}_{\beta}\, {\bar \partial }\beta
-{\pi}_{\gamma} \,{\bar \partial }\gamma
 -{\pi }_b \,{\pi }_c
-{\pi}_{\beta} \,{\pi}_{\gamma} \cr
&    -h \,( T_{\rm mat} + T_{\rm gh} )
-B \,( W_{\rm mat}+W_{\rm gh} ) +
K_{{\varphi }^i}[ c\, {\partial } {\varphi }^i + d_{ijk}\, \gamma \,\partial
{\varphi }^j \partial {\varphi }^k + a\, {\pi}_c\, \gamma \,\partial \gamma
\,\partial {\varphi }^i \cr
& + {\sqrt \hbar} ( -{\alpha }_i\, \partial c + (e_{ij}-e_{ji})\, \gamma\,
{\partial }^2 {\varphi }^j - e_{ji} \,\partial \gamma \,\partial {\varphi }^j
-a\, {\alpha }_i\, \partial ({\pi }_c \,\gamma\, \partial \gamma ) ) +
\hbar \,f_i\, {\partial }^2 \gamma ] \cr
&+ K_h \,{\pi}_b + K_B \,{\pi}_{\beta} +K_b\, {\pi}_h +K_{\beta}\, {\pi}_B
+K_{\gamma } ( c\, \partial \gamma - 2 \partial c\, \gamma ) \cr
&+ K_c [ c\, \partial c - {a\over 2} \gamma \,\partial \gamma \,\partial
{\varphi}^i \partial {\varphi }^i -a {\sqrt \hbar }\, {\alpha }_i\, \gamma\,
\partial \gamma \,{\partial }^2 {\varphi }^i +
{1\over 30 } (1-17a) \hbar
(2 \gamma\, {\partial }^3 \gamma -3 \partial \gamma\, {\partial }^2
\gamma ) ] \cr
& + K_{{\pi}_c} ( T_{\rm mat} + T_{\rm gh} ) + K_{{\pi}_{\gamma} }
( W_{\rm mat} + W_{\rm gh}). &(4.11)\cr }
$$

  Since we have already included the background charges $\alpha^i$ and the
attendant other factors of $\sqrt\hbar$ and $\hbar$ in the currents appearing
in (4.11), we shall have to expand the effective action $\Gamma$ into a
half-step series in $\hbar$:
$$
\Gamma = {\Gamma }_0 + {\hbar}^{1\over 2} {\Gamma }_{1\over 2} +
\hbar {\Gamma }_1 + {\hbar }^{3\over 2} {\Gamma }_{3\over 2} +
{\hbar }^2 {\Gamma }_2 + \cdots ,
\eqno (4.12)
$$
where $\Gamma_0$ contains all the terms in the effective action of order
$\hbar^0$. These effective-action terms in $\Gamma_0$ are just the $\hbar^0$
terms appearing in (4.11), so we may write $\Gamma_0=S_0$, where by $S_n$ we
mean the terms of order $\hbar^n$ in the extended classical action, $I_{\rm
ext}=\sum_m \hbar^{m/2}S_{m/2}$. At order
$\hbar^{\sfrac12}$, we have a similar situation,
$\Gamma_{\sfrac12}=S_{\sfrac12}$. At order $\hbar$ we encounter for the first
time loop corrections, so $\Gamma_1$ can be divided into two parts,
$\Gamma_1=S_1+\Gamma_{1,{\rm nl}}$, where $\Gamma_{1,{\rm nl}}$ contains the
(non-local) loop contributions. At order $\hbar^{\sfrac32}$,
$\Gamma_{\sfrac32}$ is obtained from one-loop corrections to $S_{\sfrac12}$.
At order $\hbar^2$, we have three different kinds of loop contributions:
one-loop contributions involving two terms from $S_{\sfrac12}$ together with
terms from $S_0$, one-loop contributions involving a term from $S_1$ together
with terms from $S_0$, and finally the true two-loop contributions, involving
only terms from $S_0$.

  As we have already stated, the gauge-fixed action (4.3) is invariant under
the transformations (4.6) at orders $\hbar^0$ and $\hbar^{\sfrac12}$, and the
transformations (4.6) are nilpotent to these same orders. Expressing these
statements in antibracket notation (2.29), we have
$$
\Big(S_0 , S_0 \Big) = 0 ; \ \ \ \ \ \ \ \ \ \ \ \ \ \
\Big(S_0 , S_{1\over 2} \Big) = 0 .
\eqno(4.13)
$$

  We must also expand the anomaly $\triangle $ in a half-step series in
$\hbar$,
$$
\triangle = \hbar {\triangle }_1
+ {\hbar}^{3\over 2} {\triangle}_{3\over 2} + {\hbar }^2
{\triangle }_2 +\cdots .
\eqno (4.14)
$$
Now we can write the anomalous Ward identity (2.21) at orders $\hbar$,
$\hbar^{\sfrac32}$ and $\hbar^2$ as the separate equations\crampest
$$
\eqalignno{
{\cal A}_1 &= {\triangle }_1 ={\delta S_0 \over \delta
{\phi }^i}\, {\delta \Gamma_1 \over \delta K_{{\phi }^i}}
+{\delta S_0 \over \delta K_{{\phi }^i}}\,
{\delta \Gamma_1 \over \delta {\phi }^i}
+{\delta S_{1\over 2 } \over \delta {\phi }^i}\,
{\delta S_{1\over 2} \over \delta K_{{\phi }^i}} ,
&(4.15a) \cr
{\cal A}_{3\over2} &= {\triangle }_{3\over 2} =
{\delta S_0 \over \delta {\phi }^i}\,
{\delta {\Gamma }_{3\over 2} \over \delta K_{{\phi }^i}}
+{\delta S_0 \over \delta K_{{\phi }^i}}\,
{\delta {\Gamma }_{3\over 2} \over \delta {\phi }^i} +
{\delta S_{1\over 2} \over
\delta {\phi }^i}\, {\delta \Gamma_1 \over \delta K_{{\phi }^i}}
+{\delta S_{1\over 2} \over \delta K_{{\phi }^i}}\,
{\delta \Gamma_1 \over \delta {\phi }^i} ,
&(4.15b) \cr
{\cal A}_2&={\triangle }_2 +{\cal A}_{2,{\rm nl}}= {\delta S_0
\over \delta{\phi }^i} \,{\delta {\Gamma }_2 \over \delta K_{{\phi }^i}} +
{\delta S_0 \over \delta
K_{{\phi }^i}}\, {\delta {\Gamma }_2 \over \delta {\phi }^i}+
{\delta S_{1\over2} \over
\delta {\phi }^i} \,{\delta {\Gamma }_{3\over 2} \over \delta K_{{\phi }^i}}
+{\delta S_{1\over2} \over \delta K_{{\phi }^i}}\,
{\delta {\Gamma }_{3\over 2} \over \delta {\phi }^i} +
{\delta \Gamma_1 \over \delta {\phi }^i} \,
{\delta \Gamma_1 \over \delta K_{{\phi
}^i}}, &(4.15c) \cr }
$$\uncramp
where ${\cal A}_n$ represents the total (local plus non-local) anomaly at order
$\hbar^n$, while ${\triangle}_n$ represents the local order-$\hbar^n$ anomaly
that remains after non-local ``dressings'' of lower-order anomalies are
separated off, in accordance with (2.21). The term ${\cal A}_{2,{\rm nl}}$ is
just such a dressing of the order-$\hbar$ anomaly.

  Using the extended Lagrangian (4.11) and the relations (3.5, 3.6),
we find the local anomalies after a straightforward but somewhat
tedious calculation:
$$
\eqalignno{ {\triangle}_1 =&
 ( {16\over 30 \pi } (1-17a) - {a\over 12 \pi } C_{\rm mat}  )
\int d^2z\, \Big(
{\gamma }\, {{\pi}_c} \,\partial {\gamma }\,
( {\partial }^3 h - {\partial }^3
{K_{{\pi}_c}} ) \cr
&\ \ \ \ \ \ \ \ \ \ \ \  +  {\partial }^3 c\,
[ {\gamma }\, K_c \,\partial \gamma - {\pi}_c \,( \partial \gamma \,B - \gamma
\,\partial B - \partial \gamma\, K_{{\pi}_{\gamma}} + \gamma \,\partial
K_{{\pi}_{\gamma}} ) ] \Big) \cr
&\ \ \ \ \ \ \ \ \ \ \ \  - {1\over 12 \pi } (100- C_{\rm mat}) \int d^2z\,
c\,( {\partial }^3 h - {\partial }^3
{K_{{\pi}_c}}) ,  &(4.16a) \cr
{\triangle}_{3\over 2} =& 0 ,   &(4.16b) \cr
{\triangle}_2 = & ( - {29\over 50 \pi } (1-17a) + {C_{\rm mat}\over 360 \pi} )
\int d^2z\, {\gamma}\,({\partial }^5 B - {\partial }^5
{K_{{\pi}_c}} ) .&(4.16c) \cr }
$$
Note that the above results contain only universal, {\it i.e.}\
purely gauge-field-dependent and ghost-dependent anomalies. We recall that this
is because we are employing in (4.11) the
renormalisations already found in Ref.\ [15] for the cancellation of
matter-dependent anomalies. We note also that the second term in
${\triangle}_1$
is same as in the Virasoro case (2.27) except for its coefficient, which is not
surprising since the Virasoro algebra is a subalgebra of the $W_3$ algebra.
Since we now have an additional pair of spin-3 ghosts, the coefficient changes
from 26 to 100.

  We next shall show that these anomalies satisfy the Wess-Zumino consistency
condition. In order to do this, we shall need parts of ${\cal A}_{2,{\rm
nl}}$ at order $\hbar^2$ that, although non-local in structure, will make local
contributions to the consistency condition when antibracketed with $S_0$:
$$\eqalign{
{\cal A}_{2,{\rm nl}}=\ft12\Big(-{29\over 50\pi}(1&-17a) + {C_{\rm
mat}\over 360 \pi} \Big) \int d^2z\,
K_{\pi_\gamma}{\partial^5\over\bar\partial}(B-K_{\pi_\gamma})\cr
&-\ft12\left(100-C_{\rm mat}\over12\pi\right) \int d^2z\,
K_{\pi_c}{\partial^3\over\bar\partial}(h-K_{\pi_c})+{\cal A}'_{2,{\rm
nl}},\cr}\eqno(4.17)
$$
where ${\cal A}'_{2,{\rm nl}}$ denotes the remaining non-local parts
of ${\cal A}_2$.

  The Wess-Zumino consistency condition for the order-$\hbar$ anomaly
${\triangle}_1$ is obtained by taking an antibracket with
$S_0$,
$$
\Big(S_0,\triangle_1\Big)=\Big(S_0,\big(S_0,\Gamma_1\big)\Big)+
\ft12\Big(S_0,\big(S_{1\over2},S_{1\over2}\big)\Big) ,\eqno(4.18)
$$
and then using the Jacobi identity for the antibracket together with the
low-order relations (4.13). Thus, we obtain the condition
$$
\Big(S_0,\triangle_1\Big)=0,\eqno(4.19)
$$
similarly to (2.30), despite the presence of the order $\hbar^{\sfrac12}$
terms in the extended Lagrangian (4.11). Insertion of the calculated anomaly
(4.16$a$) then shows after some algebra that (4.19) is indeed satisfied,
confirming the correctness of the form of (4.16$a$).

  In continuing on to check the Wess-Zumino consistency condition at order
$\hbar^2$, one needs to take into account the dressing ${\cal A}_{2,{\rm
nl}}$ of the order-$\hbar$ anomaly that is present in (4.16$c$). In
order to show how this works, we shall for simplicity consider just the local
terms appearing in the order-$\hbar^2$ consistency condition; the non-local
terms will then have to cancel separately. Taking an antibracket with
$S_0$, we have at order $\hbar^2$
$$
\Big(S_0,{\cal A}_2\Big)=\Big(S_0,\big(S_0,\Gamma_2\big)\Big) +
\Big(S_0,\big(S_{1\over2},\Gamma_{3\over2}\big)\Big) +
\Big(S_0,\big(\Gamma_1,\Gamma_1\big)\Big).\eqno(4.20)
$$
Use of the Jacobi identity, (4.13) and (4.16$b$), and restricting attention to
the local contributions in (4.20) then gives
$$
\Big(S_0,\triangle_2\Big) + \Big(S_0,{\cal A}_{2, {\rm nl}}\Big)_{\rm loc} +
\Big(S_1,\triangle_1\Big)=0,\eqno(4.21)
$$
where the second term is restricted to local contributions after taking the
antibracket. Insertion of the calculated results (4.16, 4.17) shows
that the consistency condition (4.21) indeed is satisfied, confirming
our form for the anomaly at order ${\hbar}^2$.

  The results (4.16) confirm once again that for a matter system (3.3) with
$C_{\rm mat}=100$, all the anomalies cancel [19]. Using (4.16) for a
non-critical matter system opens the way to an investigation of
anomaly-induced dynamics in the correlation functions of $W_3$ gravity along
lines generalising the work of Ref.\ [4].

\bigskip
\noindent{\bf 5. Conclusion: operator-product versus
Ward-identity anomalies}
\bigskip

  In this paper, we have reformulated the BRST quantisation procedure for
chiral worldsheet gravities by the adoption of a derivative gauge condition
and the introduction of momenta in order to put the ghost sector of the theory
back into first-order form. These simple changes to the BRST formalism for
worldsheet gravities render the formalism canonical in the sense that the BRST
transformations of all fields now arise as canonical transformations generated
by the BRST charge $Q$.

  A very simple, but apparently so far unnoticed, consequence of this
canonical structure is the following relation between the notion of an anomaly
in the BRST operator algebra, {\it i.e.}\ the failure of $Q^2$ to vanish at the
quantum level, and the anomalies (2.27, 4.16) in the BRST Ward identities.

  In the case of Virasoro gravity, interpreting the BRST charge $Q$ (2.16) as a
normal-ordered quantum operator, one may calculate $Q^2$ by standard
operator-product techniques. Writing (2.16) as the integral of a
normal-ordered operator current,
$$
Q=\oint {dz\over2\pi\im}\, J_{\rm B}(z),\eqno(5.1)
$$
where the integral in complex worldsheet coordinates is now interpreted as a
closed loop around the origin and in order to recover equivalence to the
standard mode-expansion result, one should collect the simple poles in (5.1)
using Cauchy's theorem, hence the factor of $(2\pi\im)^{-1}$ in the measure.
Calculating $Q^2$ using standard operator-product rules, one obtains
$$\eqalignno{
Q^2&\sim\hbar^2\,\oint {dz\over2\pi\im}\langle J_{\rm B}J_{\rm
B}\rangle_1(z);&(5.2a)\cr
\langle J_{\rm B}J_{\rm B}\rangle_1 &=
{1\over6}(26-D)\,c\,\partial^3c.&(5.2b)\cr}
$$
In evaluating (5.2), we have taken as usual the operator product
$J_{\rm B}(z)J_{\rm B}(w)$ and have extracted the residue of the first-order
pole $(z-w)^{-1}$ in the resulting Laurent series; the result of this
procedure is here denoted $\langle J_{\rm B}J_{\rm B}\rangle_1$.

  Equation (5.2) expresses the BRST anomaly as it is understood in conformal
field theory. Its relation to the Ward-identity anomaly (2.27) may now be
simply
stated in our reformulated BRST procedure as
$$
\hbar\, \triangle_1 = -{1\over2\pi}\int d^2z\,\Big\langle\langle J_{\rm
B}J_{\rm B}\rangle_1,
\Big({\pi}_c\, (h-K_{{\pi}_c})\Big) \Big\rangle_1.
\eqno (5.3)
$$
This can also be expressed using the gauge
fermion for the extended Virasoro-gravity action (2.23),
$$
\Psi_{\rm Vir,\ ext}={\pi}_c\, h+\sum_i(-1)^{[i]}\phi^i\,K_{\phi^i},\eqno(5.4)
$$
where $[i]$ takes the values $(0,1)$ for (bose, fermi) variables; variation of
(5.4) produces all of the non-kinetic terms in (2.23), as one can see
from (2.15) and remembering that $\delta K_{\phi^i}=0$. Using (5.4), the
relation (5.3) can be written\footnote{$^{\ddag}$}{Note, however, that the only
part of $\Psi_{\rm Vir,\ ext}$ that contributes in (5.5) is
${\pi}_c(h-K_{{\pi}_c})$.}
$$
\hbar\, \triangle_1 = -{1\over2\pi}\int d^2z\,\Big\langle\langle J_{\rm
B}J_{\rm B}\rangle_1,\Psi_{\rm Vir,\ ext}\Big\rangle_1.\eqno(5.5)
$$

  In the case of $W_3$ gravity, for the BRST operator (4.9) we find
the operator-product anomaly
$$\eqalignno{
\langle J_{\rm B}J_{\rm B}\rangle_1 &= {1\over6}(100-C_{\rm mat})\,
c\,\partial^3 c + \Big(-{16\over 15} (1-17a) + {a\over 6}
C_{\rm mat} \Big) \,\gamma\, {\pi}_c \,\partial \gamma\,
\partial^3 c\cr
&\phantom{= \pi^{-1}\oint{dz\over2\pi\im}} + \hbar\Big({29\over 25}(1-17a) -
{C_{\rm mat}\over 180}\Big) \,\gamma\, \partial^5 \gamma.&(5.6)\cr}
$$
Using this, we can express the non-vanishing Ward-identity anomalies (4.15) as
$$
\hbar\, {\triangle}_1 + \hbar^2\,{\triangle}_2 =
-{1\over 2 \pi} \int d^2z \Big\langle\langle J_B,J_B\rangle_1, \Big({\pi}_c
(h-K_{{\pi}_c})+{\pi}_{\gamma} (B-K_{{\pi}_{\gamma}})-c K_c
\Big)\Big\rangle_1,
\eqno (5.7)
$$
or, in terms of the gauge fermion for the extended $W_3$ action (4.11),
$$
\Psi_{W_3,\ {\rm ext}}=\pi_c\, h+\pi_\gamma\,
B+\sum_i(-1)^{[i]}\phi^i\, K_{\phi^i},\eqno(5.8)
$$
we can write
$$
\hbar\, {\triangle}_1 + \hbar^2\, {\triangle}_2 =
-{1\over2\pi}\int d^2z\, \Big\langle\langle J_{\rm
B}J_{\rm B}\rangle_1,\Psi_{\rm W_3,\ ext}\Big\rangle_1.\eqno(5.9)
$$
Consequently, by applying $\langle J_{\rm B}J_{\rm B}\rangle_1$ to the gauge
fermion, we have reproduced all of the local $W_3$ anomalies. Note that the
order-$\hbar^2$ local anomaly ${\triangle}_2$ arises in this relation from the
terms in
$\langle J_{\rm B}J_{\rm B}\rangle_1$ that carry
an explicit factor of $\hbar$; these
terms arise from at most double contractions, but involve the order-$\hbar$
renormalisation terms in (4.8). Thus, the local operator-product anomalies for
$W_3$ all arise from processes involving at most double contractions, in
contrast to the Feynman-diagram calculation, where genuine two-loop diagrams
are
involved.

  The simplicity of the relations (5.5, 5.9) suggests a general result for the
relation between operator-product $Q^2$ anomalies as calculated in conformal
field theory and the anomalies occurring in the BRST Ward
identities. In order to obtain this relation, it appears to be necessary to
take
care, as we have in this article, to use a legitimate gauge choice, so that the
BRST transformations of all fields arise as canonical transformations generated
by the BRST charge $Q$. It remains an interesting problem to show whether this
result is obtained in general, as well as finding its analogue in field
theories in other dimensions, such as Yang-Mills theory.
\np
\noindent{\bf Acknowledgement}
\bigskip
K.S.S. Would like to thank the Department of Physics of Texas A\&M
University and the Institute for Theoretical Physics of SUNY Stony Brook
for hospitality. Helpful discussions on the canonical formalism for chiral
$d=2$ theories with Peter Van Nieuwenhuizen and Kostas Skenderis are also
gratefully acknowledged.
\bigskip\bigskip

\centerline{\bf References }
\bigskip
\frenchspacing
\item{[1]}B. Lian and G. Zuckerman, {\sl Phys. Lett.} {\bf B254} (1991)
417;\nl
E. Witten, {\sl Nucl. Phys.} {\bf B373} (1992) 187;\nl
E. Witten and B. Zwiebach, {\sl Nucl. Phys.} {\bf B377} (1992) 55.
\item{[2]}S.K. Rama, {\sl Mod. Phys. Lett.} {\bf A6} (1991) 3531;\nl
C.N. Pope, E. Sezgin, K.S. Stelle and X.J. Wang, {\sl Phys. Lett.} {\bf B299}
(1993) 247;\nl H. Lu, C.N. Pope, X.J. Wang and K.-W. Xu,  ``The complete
cohomology of the $W_3$ string,'' preprint hep-th/9309041,
to appear in {\sl Class. Quantum Grav.}
\item{[3]}S.R. Das, A. Dhar and S.K. Rama, {\sl Mod. Phys. Lett.}
{\bf A6} (1991) 3055;\nl {\sl Int. J. Mod. Phys.} {\bf A7} (1992) 2295;\nl
A. Bilal and J.-L. Gervais, {\sl Phys. Lett.} {\bf 206B} (1988)
206;\nl {\sl Nucl. Phys.} {\bf B314} (1989)
646; {\bf B318} (1989) 576;\nl
A. Anderson, B.E.W. Nilsson, C.N. Pope and K.S. Stelle, ``The Multivalued
Free-Field Maps of Liouville and Toda Gravities,'' preprint hep-th 9401007.
\item{[4]}A.M. Polyakov,  Phys. Lett. {\bf 101B} (1981) 207; Mod. Phys.
Lett.{\bf A2} (1987) 893.
\item{[5]}E. Bergshoeff, P.S. Howe, C.N. Pope, E. Sezgin, X. Shen and K.S.
Stelle,\nl {\sl Nucl. Phys.} {\bf B363} (1991) 163.
\item{[6]}C.N. Pope, X. Shen, K.-W. Xu and K. Yuan,
{\sl Nucl. Phys.} {\bf B376} (1992) 52.
\item{[7]}K. Schoutens, A. Sevrin and P. van Nieuwenhuizen,
{\sl Nucl. Phys.} {\bf B364} (1991) 584;\nl {\bf B371} (1992) 315.
\item{[8]}M. Kato and K. Ogawa, {\sl Nucl. Phys.} {\bf B212} (1983) 443.
\item{[9]}W. Siegel and B. Zwiebach, {\sl Nucl. Phys.} {\bf B288} (1987)
332;\nl
L. Baulieu, W. Siegel and B. Zwiebach, {\sl Nucl. Phys.} {\bf B287} (1987)
93;\nl
D.Z. Freedman, J.I. Latore and K. Pilch, {\sl Nucl. Phys.} {\bf B306} (1988)
77;\nl
A. Rebhan and U. Kraemmer, {\sl Phys. Lett.} {\bf 196B} (1987) 477.
\item{[10]}C. Teitelboim, {\sl Phys. Rev.} {\bf D25} (1982) 3159.
\item{[11]}K. Schoutens, A. Sevrin and P. van Nieuwenhuizen, ``Loop
calculations in BRST quantized chiral $W_3$ gravity,'' in {\it Proc. Workshop
on
Quantum Field Theory, Statistical Mechanics, Quantum Groups and Topology (Miami
1991)}.
\item{[12]}J. Zinn-Justin, ``Renormalization of Gauge Theories,'' in
{\it Proc. Int. Summer Inst. for Theor. Physics (Bonn, 1974);}\nl
I.A. Batalin and G.A. Vilkovisky, {\sl Phys. Lett.} {\bf 102B}
(1981) 27;\nl {\sl Phys. Rev.} {\bf D28 } (1983) 2567.
\item{[13]}S. Vandoren and A. Van Proeyen, {\sl Nucl. Phys.} {\bf B411}
(1994) 257.
\item{[14]}V. Del Duca, L. Magnea and P. Van Nieuwenhuizen, {\sl Int. J.
Mod. Phys.} {\bf A3} (1988) 1081.
\item{[15]}C.N. Pope, L.J. Romans and K.S. Stelle, {\sl Phys. Lett.} {\bf
268B} (1991) 167.
\item{[16]}B. Zumino, in {\it Supersymmetry and applications:
superstrings, anomalies and supergravity,} eds. G.W. Gibbons, S.W. Hawking and
P.K. Townsend\nl (Cambridge University Press 1986);\nl
P.S. Howe, U. Lindstrom and P. White, {\sl Phys. Lett.} {\bf 246B } (1990)
430;\nl
F. Bastianelli, {\sl Phys. Lett.} {\bf 263B } (1991) 411.
\item{[17]}A.B. Zamolodchikov, {\sl Teo. Mat. Fiz.} {\bf 65} (1985) 644.
\item{[18]}L.J. Romans, {\sl Nucl. Phys.} {\bf B352} (1991) 829.
\item{[19]}J. Thierry-Mieg, {\sl Phys. Lett.} {\bf 197B } (1987) 368.
\item{[20]}C.M. Hull, {\sl Phys. Lett.} {\bf 240B} (1989) 110.
\item{[21]}R.Kallosh, {\sl Nucl. Phys.} {\bf B141} (1978) 141;\nl
G. Sterman, P.K. Townsend and P. van Nieuwenhuizen, {\sl Phys. Rev.}
{\bf D17} (1978) 1501;\nl
K.S. Stelle and P.C. West, {\sl Phys. Lett.} B74 (1978) 330.
\bye